\documentclass[10pt]{iopart}
\usepackage{graphicx}
\usepackage{balance}
\usepackage{color}
\usepackage{cite}
\expandafter\let\csname equation*\endcsname\relax
\expandafter\let\csname endequation*\endcsname\relax
\usepackage{amsmath}
\usepackage{hyperref}

\hypersetup{
    colorlinks=true,
    linkcolor=black,
    citecolor=black,
    filecolor=black,
    urlcolor=black,
}

\begin{document}

\title[Wave control through soft microstructural curling]{Wave control through soft microstructural curling: bandgap shifting, reconfigurable anisotropy and switchable chirality}

\author{Paolo Celli$^{1}$, Stefano Gonella$^{1}$, Vahid Tajeddini$^{2}$, Anastasia Muliana$^{2}$, Saad Ahmed$^{3}$, Zoubeida Ounaies$^{3}$}

\address{$^{1}$ Department of Civil, Environmental, and Geo- Engineering, University of Minnesota, Minneapolis, MN 55455, USA}
\address{$^{2}$ Department of Mechanical Engineering, Texas A\&M University, College Station, TX 77840, USA}
\address{$^{3}$ Department of Mechanical and Nuclear Engineering, The Pennsylvania State
University, University Park, PA 16802, USA}
\ead{sgonella@umn.edu}
\vspace{10pt}
\begin{indented}
\item[]{{\color{red}\underline{\bf Published article}}: \emph{Smart Materials and Structures} {\bf 26} (3), 035001 (2017);\quad \url{http://dx.doi.org/10.1088/1361-665X/aa59ea}}
\end{indented}

\begin{abstract}
In this work, we discuss and numerically validate a strategy to attain reversible macroscopic changes in the wave propagation characteristics of cellular metamaterials with soft microstructures. The proposed cellular architecture is characterized by unit cells featuring auxiliary populations of symmetrically-distributed smart cantilevers stemming from the nodal locations. Through an external stimulus (the application of an electric field), we induce extreme, localized, reversible \emph{curling} deformation of the cantilevers---a shape modification which does not affect the overall shape, stiffness and load bearing capability of the structure. By carefully engineering the spatial pattern of straight (non activated) and curled (activated) cantilevers, we can induce several profound modifications of the phononic characteristics of the structure: generation and/or shifting of total and partial bandgaps, cell symmetry relaxation (which implies reconfigurable wave beaming), and chirality switching. While in this work we discuss the specific case of composite cantilevers with a PDMS core and active layers of electrostrictive terpolymer P(VDF-TrFE-CTFE), the strategy can be extended to other smart materials (such as dielectric elastomers or shape-memory polymers).
\end{abstract}

%
\vspace{2pc}
\ioptwocol
%

\section{Introduction} \label{sec:intro}
Metamaterials are media whose internal architecture is carefully designed to elicit properties that are not attainable by conventional materials available in nature. Some of the most promising opportunities for metamaterials engineering are found in the control of acousto-elastic waves; \emph{acoustic metamaterials} are typically (but not necessarily) periodic structures which display exotic wave manipulation effects, such as sub-wavelength filtering~\cite{Liu_SCIENCE_2000} and waveguiding~\cite{Lemoult_NAT-PHYS_2013}, negative refraction~\cite{Zhu_NAT-COMM_2014} and cloaking~\cite{Torrent_NJoP_2008}. One of the main drawbacks of metamaterials is their inherent \emph{passivity}: a metamaterial is typically designed for a specific task and for a prescribed operational condition. Therefore, its geometry is determined and  ``frozen'' at the design stage. A modification of the excitation scenario (e.g., a variable frequency content) would most likely result in the need for a full re-design of the medium. To alleviate this issue, several strategies for active or adaptive control have been introduced over the past decade. The simplest approach involves a manual adjustment of the position of the constitutive elements of a periodic medium~\cite{Goffaux_PRB_2001, Romero-Garcia_JPD_2013}. More automated strategies revolve around the application of external non-mechanical stimuli to modify the mechanical properties of smart material inserts (e.g., shape memory alloys~\cite{Ruzzene_JVA_2000}, shunted piezoelectric~\cite{Thorp_SMS_2001, Airoldi_NJoP_2011, Wang_SMS_2011, Casadei_JAP_2012, Bergamini_ADMA_2014, Zhu_APL_2016, Yi_SMS_2016, Cardella_SMS_2016}, magnetoelastic~\cite{Robillard_APL_2009, Schaeffer_JAP_2015, Allein_APL_2016, Wang_ADVMAT_2016}, electrorheological~\cite{Yeh_PHYB_2007} and magnetorheological~\cite{Xu_SSC_2013} phases, thermally-stimulable polymeric materials~\cite{Walker_APL_2014}, aerodynamic-loading-sensitive elements~\cite{Casadei_JSV_2014}), eventually resulting in a reversible modification of the wave propagation characteristics of the metamaterial. Tunable phononic characteristics are also attainable using nonlinear periodic structures. For example, by inducing buckling in some structural elements via the application of static loads, it is possible to trigger large deformations that can result in dramatic pattern reconfigurations~\cite{Bertoldi_PRB_2008, Wang_PRL_2014, Maurin_JSV_2014, Rudykh_PRL_2014}. Other avenues leverage the onset of geometric and material nonlinearities. For example, changing the level of precompression of a granular crystal results in a variation of its nonlinear characteristics~\cite{Daraio_PRE_2005}; this concept has been exploited to produce media with tunable bandgaps~\cite{Narisetti_JVA_2010}, wave directivity~\cite{Narisetti_JVA_2011} and tunable wave focusing capabilities~\cite{Spadoni_PNAS_2010}. Finally, nonlinearly-induced higher-harmonics have recently been used to produce wavefields displaying modal mixing and augmented directivity patterns~\cite{Ganesh_PRL_2015}.

Recent advances in the field of soft active materials have opened new opportunities for tunability~\cite{Yang_SMS_2008, Gei_IEEE_2011, Bayat_JVA_2015, Nouh_JIMSS_2015, Jia_SMS_2016, Galich_IJSS_2016, Getz_IJSS_2016}. Pioneering, in this sense, are the works of Yang and Chen~\cite{Yang_SMS_2008}, who first proposed dielectric elastomers to tune the wave characteristics of periodic structures, of Bayat and Gordaninejad~\cite{Bayat_JVA_2015}, who studied magnetorheological shape transforming lattices, and of Nouh et al.~\cite{Nouh_JIMSS_2015}, who investigated periodic plates with soft electrically-stiffened PVDF inclusions. While these works introduce soft active materials in the context of wave control devices, they only partially take advantage of the dramatic shape modifications enabled by soft active materials, such as dielectric elastomers~\cite{Ahmed_SMS_2014, Shian_AM_2015} and shape memory polymers~\cite{Felton_SOFT_2013, Ge_SMS_2014}. In this work, in light of these recent advances, we discuss the effects of \emph{localized} and \emph{reversible} shape modifications on the spectro-spatial wave control characteristics of soft cellular structures. The shape modifications are internal and \emph{localized}, since they occur at the level of an \emph{auxiliary microstructure}---here comprising non-structural cantilever elements. The \emph{reversibility} comes from the fact that the cantilevers are equipped with smart actuators (here we consider layers of P(VDF-TrFE-CTFE) electrostrictive terpolymer~\cite{Sigamani_SMASIS_2014} on a PDMS substrate). Excited with an external (electrical, in this case) stimulus, the actuators cause the auxiliary cantilevers to \emph{curl} (i.e., experience extreme---yet reversible---rolling deformation), thus modifying their dynamic properties and their effect on the global wave manipulation capabilities of the medium. The idea introduced in this work is a logical continuation of a paradigm that has recently gained traction in the arena of phononic crystals/metamaterials~\cite{Celli_JAP_2014, Kroedel_AFM_2014, Celli_APL_2015}, based on the idea of decoupling the static (load-bearing) properties from its dynamic (wave control) functionalities. Thanks to the dramatic shape modifications that can be achieved, the strategy presented herein is expected to produce more pronounced effects than previous implementations, which merely relied on material property correction. As a first step, we briefly discuss how global microstructural curling produces reversible shifts in the bandgaps landscape. We note that this initial part of our analysis is conceptually similar to a recent investigation independently pursued by Zhang et al~\cite{Zhang_SCIREP_2016} in the context of 3D printed shape memory polymers, in which the authors briefly touch on the appearance/disappearance of phononic bandgaps. In this respect, we hereby attempt to offer a complete rationale of the physical mechanisms that govern the evolution of the bandgap behavior induced by microstructural shape changes. We then proceed to show how selective curling can lead to \emph{symmetry relaxation} of the unit cell, which in turn produces partial bandgaps which give way to focused wave patterns (reconfigurable anisotropy). Finally, we discuss how these internal shape modifications can introduce non-trivial chiral effects in the response of the periodic medium.

The paper is organized as follows. In Sec.~\ref{sec:str} we introduce our strategy for reconfigurable tunability of the lattice characteristics. In Sec.~\ref{sec:curl} we discuss how we model the curling of a cantilever beam with smart material inserts in response to an applied electric field. In Sec.~\ref{sec:wave} we report the results of the wave analysis, in terms of bandgap shifting, reconfigurable anisotropy and switchable chirality. Finally, the conclusions of our work are drawn in Sec.~\ref{sec:con}.

\section{Architecture and curling strategy}
\label{sec:str}
\begin{figure} [!htb]
\centering
\includegraphics[scale=1.38]{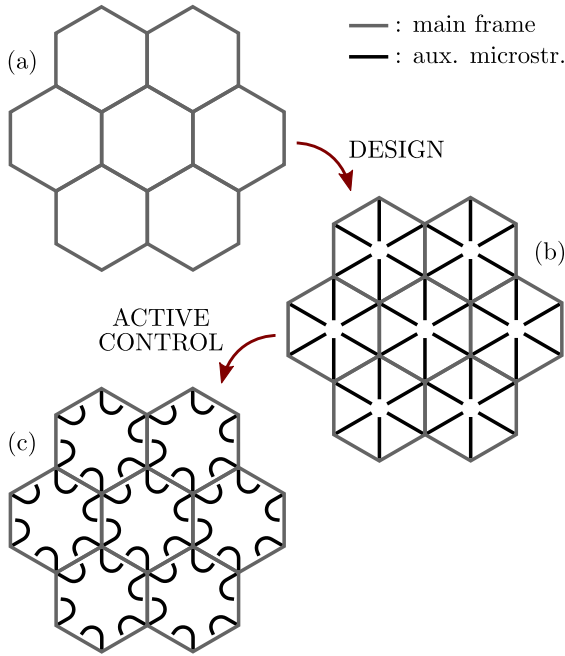}
\caption{Wave control strategy at a glance. (a) Regular hexagonal (RH) lattice frame. (b) RH frame with an auxiliary microstructure (straight cantilevers). (c) Example of architecture attainable through a smart-material-induced shape transformation (curling) of the microstructure.}
\label{fig:idea}
\end{figure}

Given a cellular architecture, e.g. the regular hexagonal (RH) lattice sketched in Fig.~\ref{fig:idea}a, we introduce a symmetric population of microstructural elements, consisting of cantilevers located at the nodal locations, as shown in Fig.~\ref{fig:idea}b.
These internal elements constitute an \emph{auxiliary microstructure}, in that their addition (or removal) does not affect the connectivity of the \emph{primary lattice} (which is still regular hexagonal) and therefore does not affect its static properties nor its load bearing capabilities. The auxiliary cantilevers behave, to all intents and purposes, as resonators capable of enriching the dispersion relation (wave response) of the cellular medium by opening \emph{locally resonant bandgaps}. Expanding on this passive architecture, we now let the cantilevers be instrumented with layers of active material and we apply external stimuli to produce reversible changes in their local shape, as sketched in Fig.~\ref{fig:idea}c. These shape transformations, which are localized at the microstructural level, cause a modification of the resonant frequency of the cantilevers and, in turn, a modification the global dynamic response (altering the frequency location of the locally resonant bandgaps). The net result is a medium with tunable wave characteristics where, remarkably, the tunability strategy does not interfere with other important structural functionalities.

\section{Nonlinear deformation of soft, smart composite cantilevers}
\label{sec:curl}
\begin{figure} [!htb]
\centering
\includegraphics[scale=1.38]{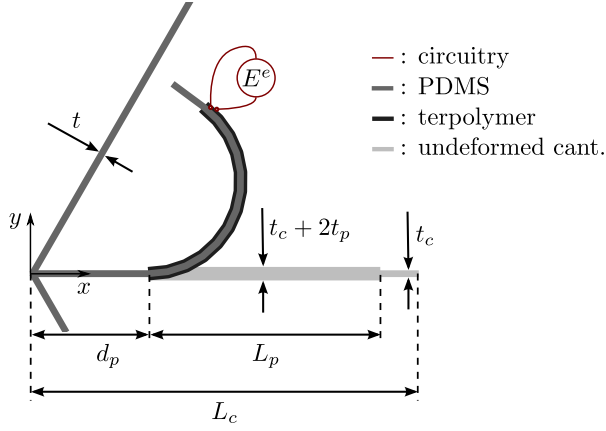}
\caption{Counterclockwise curling of a terpolymer-PDMS composite cantilever beam, obtained by activating the bottom patch with an electric field $E^e=150\,\mathrm{MV m^{-1}}$.}
\label{fig:curling}
\end{figure}

To obtain the levels of curling necessary to induce macroscopic changes in the response of the cantilevers, we are required to work with a compliant material substrate. To this end, we consider a cellular solid skeleton (main lattice plus cantilevers) made of Polydimethylsiloxane (PDMS), a soft polymer. Each cantilever is instrumented with two thin patches of a soft active material, as shown in Fig.~\ref{fig:curling}. We choose to work with P(VDF-TrFE- CTFE), an electrostrictive terpolymer capable of inducing $\sim 4\%$ electrostrictive strain. This high electrostrictive strain coupled with its high Young’s modulus makes this polymer uniquely suited for this internal cantilever application~\cite{Sigamani_SMASIS_2014, Madden_IEEE_2004}. Note that, due to the nature of the electrostriction phenomenon, the two patches are only capable of contracting in the thickness direction and expanding in the planar direction in response to a through-the-thickness electric field. For this reason, we only activate one patch at a time: activating the bottom patch allows for counterclockwise curling, while activating the top one causes the cantilever to curl clockwise. The mechanical properties for the PDMS substrate are selected within the range of properties that can be achieved with off-the-shelf PDMS kits: Young's modulus $E=2\,\mathrm{MPa}$, Poisson's ratio $\nu=0.5$, density $\rho=965\,\mathrm{kg\,m^{-3}}$. The properties of P(VDF-TrFE-CTFE) are: Young's modulus $E_p=200\,\mathrm{MPa}$, Poisson's ratio $\nu_p=0.48$, density $\rho_p=1300\,\mathrm{kg\,m^{-3}}$, coefficient of electrostriction $\beta=\beta_{13}=3\cdot10^{-18}\,\mathrm{m^2V^{-2}}$, breakdown electric field $E^e_b= 350\,\mathrm{MV\,m^{-1}}$.

A single cantilever (before and after the application of an electric field to the bottom patch), with all its characteristic dimensions, is shown in Fig.~\ref{fig:curling}.
The two lines oriented at $\pm60\mathrm{^o}$ and stemming from the origin represent portions of the lattice links belonging to the hexagonal lattice structure. Throughout this analysis, we consider the following dimensions (which have been carefully selected to be compatible with fabrication methods currently available): the length of each lattice link is $L=2.5\,\mathrm{cm}$, the thickness of a lattice link is $t=L/50=500\,\mathrm{\mu m}$, the length of a cantilever is $L_c=0.8\cdot L=2\,\mathrm{cm}$, the thickness of a cantilever is $t_c=L_c/65=308\,\mathrm{\mu m}$, the distance from the base of the cantilever to the beginning of the patch is $d_p=0.3\cdot L_c=6\,\mathrm{mm}$, the length of a patch is $L_p=0.6\cdot L_c=12\,\mathrm{mm}$, the thickness of a patch is $t_p=10\,\mathrm{\mu m}$.

To predict the curling of the terpolymer-PDMS sandwich cantilevers, we resort to an analytical model largely inspired by the work of Tajeddini and Muliana~\cite{Tajeddini_COMPSTRUCT_2015}---here adapted to the special case in which only one patch at a time is activated{\color{blue}~\cite{Wang_JIMSS_1991}}. The key aspects of the model are summarized in the following (for a complete and detailed account of the formulation, refer to the Supplementary Data (SD) Section). The kinematics are based on Reissner's finite-strain beam theory~\cite{Reissner_ZAMP_1972}, with the additional assumptions of initially straight configuration and shear indeformability~\cite{Irschik_ACTAMECH_2009, Muliana_IJNLM_2015}; the latter assumption restricts the model to the treatment of slender beams. The nonlinear strain-displacement relations are:
\begin{equation}
\frac{du}{dx}=(1+\epsilon_{x0})\cos{\phi}-1\,\,,
\label{eq:k1M}
\end{equation}
\begin{equation}
\frac{dv}{dx}=(1+\epsilon_{x0})\sin{\phi}\,\,,
\label{eq:k2M}
\end{equation}
where $u$ and $v$ are the beam's axial and lateral displacements, $\epsilon_{x0}$ is the axial strain at the beam's neutral axis and $\phi$ is the rotational angle of the cross section of the deformed beam.

The composite cantilever is equipped with thin patches of a soft (compliant) active material, perfectly bonded to the substrate and symmetrically placed with respect to the neutral axis of the beam. In response to an external stimulus, a patch exerts axial forces on the substrate at their interface. Due to the fact that, for the considered geometry, the axial strain at the centroidal axis ($\epsilon_{x0}$) is typically negligible, the action of the patch is effectively akin to a constant bending moment applied to the span of the substrate sandwiched between the patches. In light of these observations, the following relations can be written:
\begin{equation}
\epsilon_{x0}=0\,\,,
\label{eq:epsM}
\end{equation}
\begin{equation}
\frac{d\phi}{dx}=\frac{M_{ac}}{EI_c}\,\,,
\label{eq:phiM}
\end{equation}
where $I_c$ is the second moment of area of the substrate ($I_c=b\,t_c^3/12$), and $M_{ac}$ is the moment due to the actuator's action. Depending on whether we are activating the bottom (B) or top (T) patch, the moment can assume different values. If we only activate the bottom patch, the moment applied to the substrate is:
\begin{equation}
M_{ac}=\frac{E_p\,E\,b\,t_p\,t_c^2}{2(t_c\,E+6t_p\,E_p)}\,\epsilon_p\,\,;
\label{eq:MacBM}
\end{equation}
if we only activate the top patch, the resulting moment is:
\begin{equation}
M_{ac}=-\frac{E_p\,E\,b\,t_p\,t_c^2}{2(t_c\,E+6t_p\,E_p)}\,\epsilon_p\,\,,
\label{eq:MacTM}
\end{equation}
where $\epsilon_p$ is the free strain of the patch (strain undergone by an unconstrained patch under the action of the electric field); for our specific choice of soft active material, i.e., for an electrostrictive terpolymer, the free strain is:
\begin{equation}
\epsilon_p=\,\beta\,({E^e})^2\,\,,
\end{equation}
where $E^e$ is the applied electric field. Note that, to derive Eqs.~\ref{eq:MacBM}-\ref{eq:MacTM}, we also assumed the patches to be much thinner than the substrate; as a consequence, we can safely assume the axial stress to be constant along the patch thickness. It is important to note that, as a result of the slenderness and of the pure-bending actuation, the established deformation field features large rotations and small strains.

\begin{figure*} [!htb]
\centering
\includegraphics[scale=1.38]{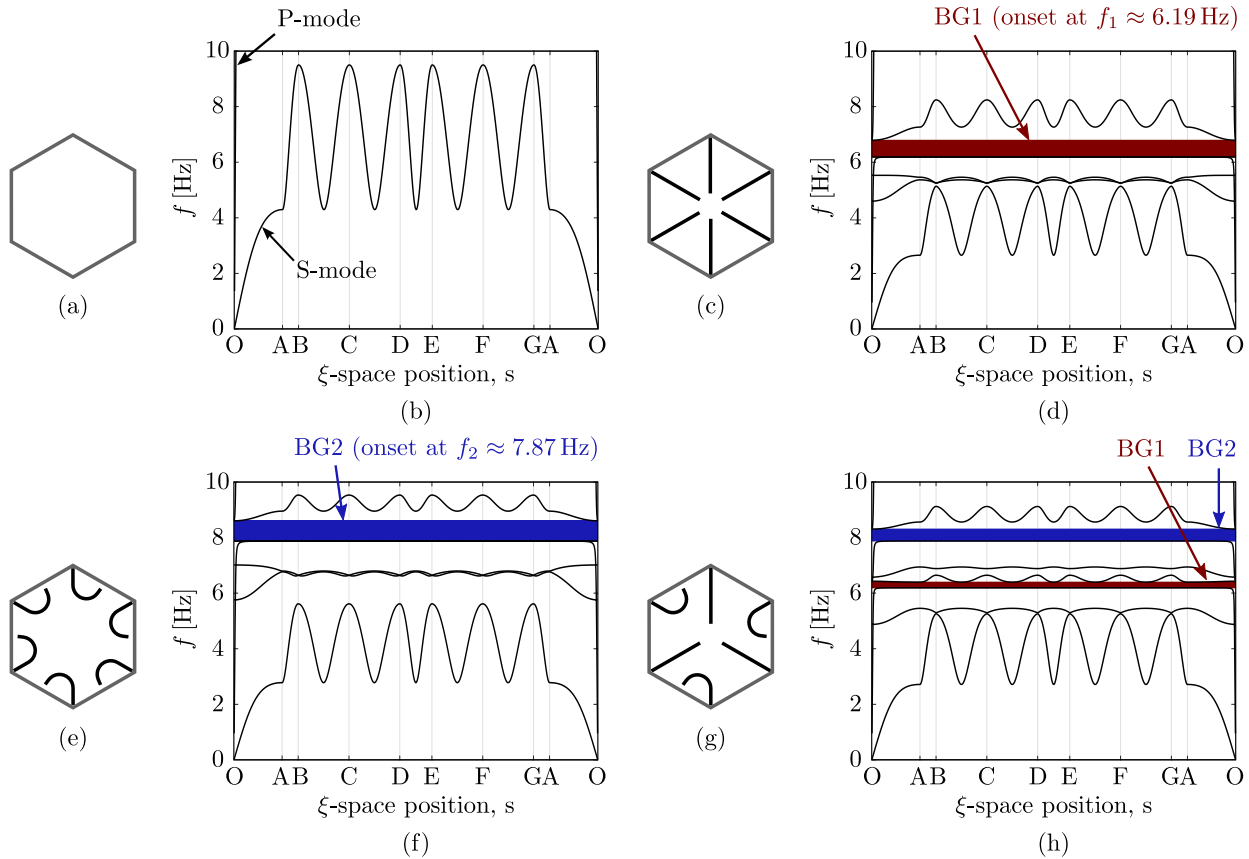}
\caption{Unit cell analysis perspective on bandgap shifting. (a) Regular hexagonal architecture. (b) Low-frequency region of the band diagram for the cell in (a). Note the steep P-mode running almost parallel to the vertical axis. (c) Microstructured architecture with a symmetric population of straight cantilevers. (d) Band diagram for the cell in (c). (e) Architecture with all curled cantilevers, resulting from the application of a $150\,\mathrm{MV m^{-1}}$ electric field to the bottom patch of each cantilever. (f) Band diagram for the cell in (e). (g) Hybrid architecture displaying three straight and three curled cantilevers. (h) Band diagram for the cell in (g).}
\label{fig:bgap}
\end{figure*}

The boundary value problem is solved in a piecewise fashion. Given the geometry considered in Fig.~\ref{fig:curling}, we can identify three intervals: $0\leq x\leq d_p$, where $M_{ac}=0$ due to the absence of patches, $d_p\leq x\leq L_p+d_p$, where $M_{ac}$ is given by Eq.~\ref{eq:MacBM} or Eq.~\ref{eq:MacTM} (depending on whether we are activating the bottom or top patch, respectively), and $L_p+d_p\leq x\leq L$ where again $M_{ac}=0$. The piecewise solution is given below. For $0\leq x\leq d_p$:
\begin{equation}
\left \{\!\!\!\begin{array}{l}
u(x)=0\\
v(x)=0 \end{array}\,\,\,\right.
\label{eq:s1}
\end{equation}
For $d_p\leq x\leq d_p+L_p$:
\begin{equation}
\left \{\!\!\!\begin{array}{l}
u(x)=\frac{EI_c}{M_{ac}}\sin{\frac{M_{ac}(x-d_p)}{EI_c}} -x+d_p\\
v(x)=\frac{EI_c}{M_{ac}}\left[ 1-\cos{\frac{M_{ac}(x-d_p)}{EI_c}} \right] \end{array}\,\,\,\right.
\label{eq:s2}
\end{equation}
For $d_p+L_p\leq x\leq L$:
\begin{equation}
\left \{\!\!\!\begin{array}{l}
u(x)\!=\!\cos\!\frac{M_{ac}L_p}{EI_c}(x\!-\!d_p\!-\!L_p)\!-\!x\!+\!d_p\!+\!\frac{EI_c}{M_{ac}}\sin\!{\frac{M_{ac}L_p}{EI_c}}\\
v(x)\!=\!\sin\!\frac{M_{ac}L_p}{EI_c}(x\!-\!d_p\!-\!L_p)\!+\!\frac{EI_c}{M_{ac}}\!\left(1\!-\!\cos\!{\frac{M_{ac}L_p}{EI_c}}\right)\end{array}\right.
\label{eq:s3}
\end{equation}
The curled shape shown in Fig.~\ref{fig:curling} is obtained by implementing Eqs.~\ref{eq:s1}-\ref{eq:s3} with the selected geometric and material parameters, for an imposed electric field $E^e=150\,\mathrm{MV m^{-1}}$ applied to the bottom patch only.

\section{Phononic analysis}
\label{sec:wave}
The analysis of the phononic characteristics of the soft hexagonal lattice with straight/curled microstructure is carried out using a unit cell discretized with beam elements, and in-house MATLAB routines. Throughout this work, we refer to the cantilevers as \emph{straight} or \emph{curled}, according to whether they are in the undeformed or deformed configuration. The finite element model of the curled cantilevers is obtained by discretizing the curved profile predicted by the analytical model discussed in Sec.~\ref{sec:curl} (shown in Fig.~\ref{fig:curling}). In addition to accounting for the effect of the patches in the form of equivalent bending moment loads, we also explicitly incorporate the influence of the terpolymer patches on the mass and stiffness matrices by taking into account the multi-material structure of the cantilevers' layered cross section according to the relative thickness of the layers. While a nonlinear model is employed to predict the curling deformation of the cantilevers, a linear small-on-large analysis suffices to describe the superimposed small deformations experienced during a transient wave event, which are a perturbation about the equilibrium deformed state reached after curling. For more details on this, including a numerical justification for the validity of the small-on-large model assumption, refer to the SD Section. It is of course implied that the static moments that cause the cantilevers to curl are preserved throughout the dynamic analysis. Also note that, throughout our analysis, we neglect viscoelastic damping effects, which are outside the proof-of-concept scope of this work, although we recognize that they may significantly affect the predicted wave manipulation effects in future experimental settings.

\subsection{Bandgap shifting}
\label{sec:bg}
The most intuitive manifestation of microstructural shape changes is the onset of bandgap tunability. This effect is here illustrated by comparing the wave response of several unit cell configurations---with straight cantilevers, curled cantilevers, or a combination of both---at different frequencies. This result is achieved through a classical Bloch analysis which yields the dispersion relation for the lattice~\cite{Phani_JASA_2006}. Dispersion relations are here portrayed in the form of \emph{band diagrams} (Fig.~\ref{fig:bgap}), obtained by plotting the frequency values computed at points of the reciprocal wave vector space marked by a coordinate s running along the contour of the First Brillouin zone (BZ) of the lattice~\cite{Brillouin}. Details on the BZ, including the significance of points O, A, ..., G are reported in the SD Section. Note that, in order to guarantee that the adopted representation is automatically compatible with the entire spectrum of symmetries (and lack thereof) that are established by arbitrarily curling the microstructural elements, we here work with the entire First Brillouin zone instead of the Irreducible Brillouin zone (IBZ) of the hexagonal lattice, which would only be applicable in selected highly-symmetric cases. More information on this point is also reported in the SD Section. This selection results in more complex (at times redundant, yet always complete) band diagrams containing a wealth of information on wave directivity. Details on how to navigate these plots and extract the necessary phononic characteristics are provided below.

\begin{figure} [!htb]
\centering
\includegraphics[scale=1.38]{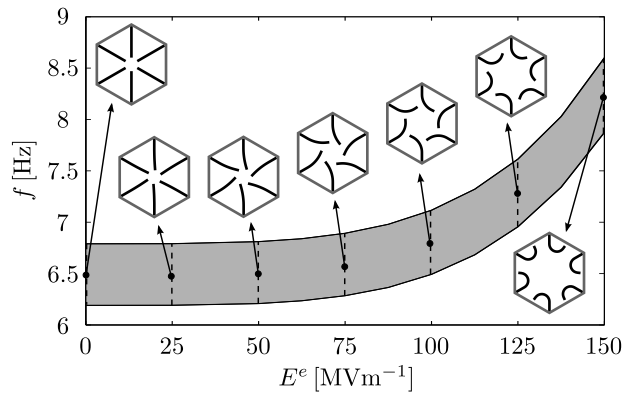}
\caption{Bandgap evolution as a function of the electric field applied to the bottom patch of each cantilever ($E^e$). For a certain value of $E^e$, the bandgap spans the frequency range encompassed by the gray band. The inserts represent the unit cell at selected values of the electric field.}
\label{fig:bgmap}
\end{figure}

We begin our analysis by considering the reference case of a regular hexagonal lattice without cantilevers. The unit cell and its band diagram are shown in Figs.~\ref{fig:bgap}a-b. In the frequency range of interest, we observe two modes of wave propagation: a slower S-mode, known to be dominated by beam bending deformation, and a very fast P-mode (note the almost vertical steep slope), dominated by longitudinal beam deformation mechanisms. In Figs.~\ref{fig:bgap}c-d we can see that adding a population of straight cantilevers yields a \emph{locally resonant bandgap} (highlighted in red and labeled BG1), along with the appearance of localized new modes which originate from the ``splitting'' of the S-mode and cluster around the bandgap. Note that the onset of the bandgap ($f_1\approx6.19\,\mathrm{Hz}$) coincides with the first natural frequency of one of the straight composite cantilevers. When computing the band diagrams for the unit cells with cantilevers, we experienced some numerical issues when s coincides with O (i.e., with the origin of the $k$-plane); details on this issue and on the reason why it does not affect our results are given in the SD Section. Application of an electric field to the bottom patch of each cantilever ($E^e=150\,\mathrm{MV\,m^{-1}}$) forces initially-straight cantilevers to curl counterclockwise, resulting in the architecture shown in Fig.~\ref{fig:bgap}e. Intuition suggests that the curled cantilevers are stiffer than their straight counterparts, thus behaving as frequency-upshifted resonators; indeed, the locally resonant bandgap (now highlighted in blue and labeled BG2) shifts upwards. Also, the onset of the bandgap ($f_2\approx7.87\,\mathrm{Hz}$) coincides now with the first natural frequency of a curled cantilever. The shift can be quantified by evaluating the relative change between the onsets of the two bandgaps, and it amounts to $27\%$ for our choice of electric field. To understand how the bandgap gradually evolves as a function of the electric field, see Fig.~\ref{fig:bgmap}.
As we increase the electric field from $0\,\mathrm{MV\,m^{-1}}$ up to $150\,\mathrm{MV\,m^{-1}}$ and the cantilevers are progressively curled and stiffened, the bandgap shifts towards higher frequencies.
Finally, we explore a hybrid scenario where three cantilevers of the unit cell are curled while the others are undeformed, as shown in Fig.~\ref{fig:bgap}g. In this case, we observe bandgaps associated with both types of resonators, with the onsets of BG1 and BG2 coinciding with those observed in Fig.~\ref{fig:bgap}d and Fig.~\ref{fig:bgap}f.

\begin{figure*} [!htb]
\centering
\includegraphics[scale=1.38]{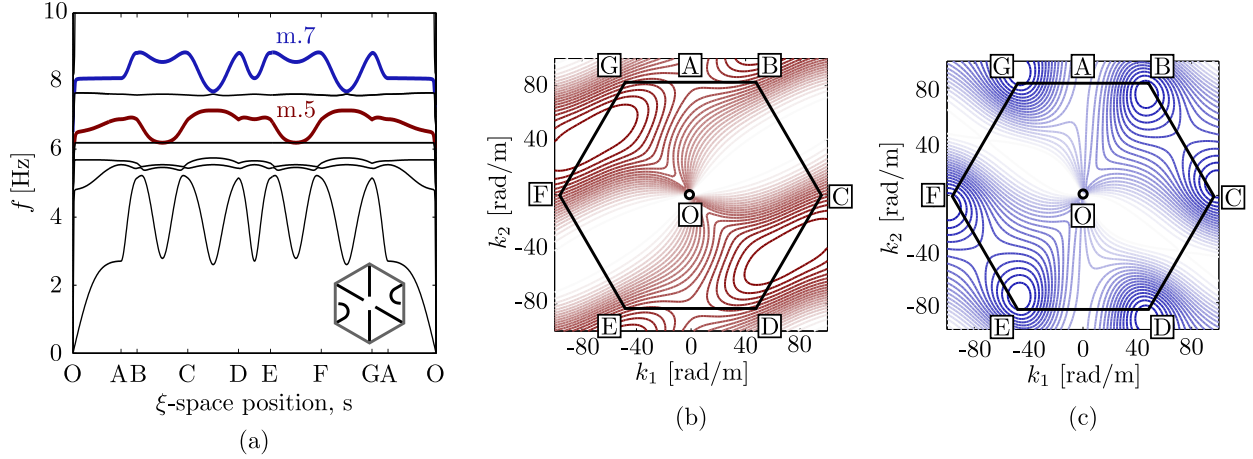}
\caption{Reconfigurable anisotropy, from a unit cell analysis perspective. (a) Band diagram for an architecture with relaxed cell symmetry; the unit cell is also sketched as an insert in (a). (b) Cartesian iso-frequency contour of the 5th mode of the dispersion relation, labeled ``m.5'' in (a). (c) Cartesian iso-frequency contour of the 7th mode of the dispersion relation, labeled ``m.7'' in (a). The hexagonal contour in (b) and (c) is the contour of the First Brillouin zone (BZ): characteristic points of the BZ are properly labeled.}
\label{fig:diruc}
\end{figure*}
\begin{figure*} [!htb]
\centering
\includegraphics[scale=1.38]{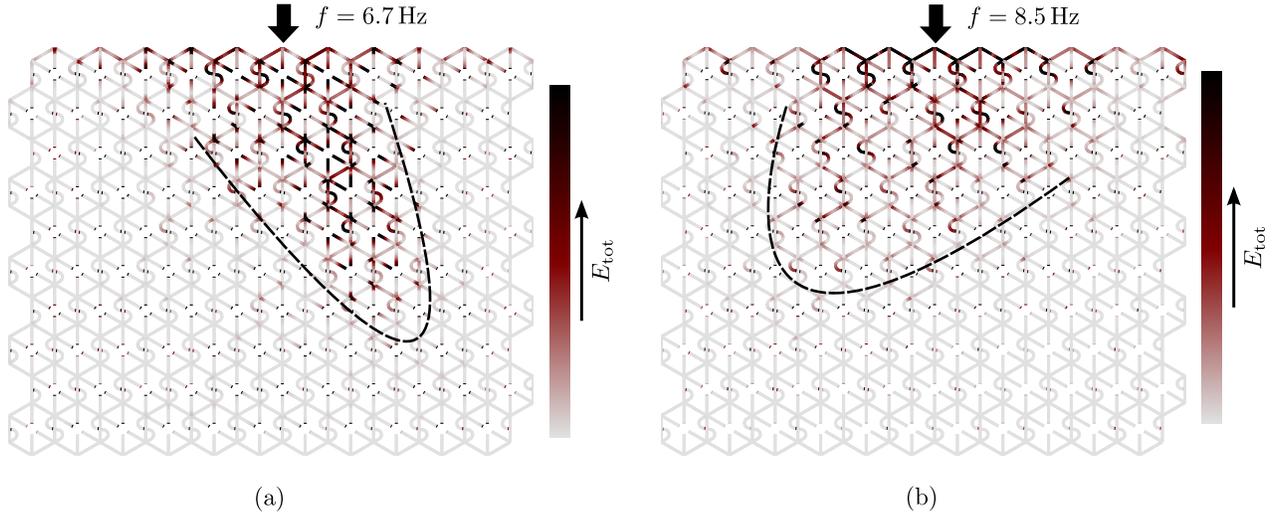}
\caption{Reconfigurable anisotropy captured via full scale simulations. Both wavefields represent the response of a finite lattice characterized by the same relaxed cell symmetry architecture shown in Fig.~\ref{fig:diruc}a and comprising $11\times10$ unit cells. The wavefields, excited by a 7-cycle burst signal, correspond to the total mechanical energy in the structure at a given time instant. The dashed lines provide a guide to the eye, and are meant to loosely bound the high-energy regions. (a) Response to a burst with carrier frequency $6.7\,\mathrm{Hz}$. (b) Response to a burst with carrier frequency $8.5\,\mathrm{Hz}$.}
\label{fig:dirfss}
\end{figure*}

\subsection{Reconfigurable anisotropy}
\label{sec:anis}
In this section, we discuss how, through a strategy that allows individual resonator tuning and asymmetric microstructural reconfiguration, we can achieve profound spatial wave manipulation effects. To this end, we revisit and enhance the idea of \emph{relaxed cell symmetry}~\cite{Celli_JAP_2014, Celli_APL_2015}, which we had previously introduced in the context of piezo-shunt-controlled lattices. Recall that the anisotropic wave patterns that are established in periodic structures~\cite{Langley_JSV_1996} usually reflect closely the symmetry (or lack thereof) of the unit cell. We can therefore expect that, by curling selected subsets of cantilevers, we would alter the symmetry landscape of the cell, thus inducing non-symmetric anisotropy patterns with pronounced wave beaming characteristics (in contrast with the symmetric behavior of the host lattice). This scenario is explored in Fig.~\ref{fig:diruc}. We can see that two of the cantilevers have been activated and curled, while the others are left in their undeformed state. From the band diagram in Fig.~\ref{fig:diruc}a, we can see that the loss of symmetry of the unit cell has drastic repercussions on its response. This is especially visible for the $5^{\mathrm{th}}$ and $7^{\mathrm{th}}$ modes (highlighted in red and blue, respectively), and it can be best appreciated by looking at the corresponding iso-frequency contours in Figs.~\ref{fig:diruc}b-c (i.e., the dispersion surface of the selected mode is sliced at different frequencies; an increase in frequency is associated with a transition from light to dark contours). The hexagon bounds the region of the Cartesian wave vector plane corresponding to the BZ, and the points highlighted on the contour are the same points indicated on the abscissa of Fig.~\ref{fig:diruc}a. We can see that both the $5^{\mathrm{th}}$ and the $7^{\mathrm{th}}$ mode lose the typical six-folded symmetry of hexagonal lattices. Taking a closer look at the $5^{\mathrm{th}}$ branch, we notice a \emph{partial bandgap} manifesting as a pair of dips in the branch, spanning the BC and EF edges of the BZ. The presence of the partial bandgap and the morphology of the dispersion surface suggest that the wave response should be attenuated along directions characterized by wavevectors pointing from O to points on the BC and EF edges, and, conversely, be focused along directions OD and OG (and neighboring ones). Similar considerations can be made for the $7^{\mathrm{th}}$ mode, which features a partial bandgap spanning the CD and FG edges, with wave beaming expected along directions OE and OB.

To validate the predictions from the unit cell analysis, we test the response of a finite lattice comprising $11\times10$ unit cells (Fig.~\ref{fig:dirfss}) with the same architecture shown in Fig.~\ref{fig:diruc}a (simulations are performed with a Newmark-$\beta$ time-integration algorithm).
The bottom nodes of the lattice are clamped, and an in-plane excitation is applied to the mid-point of the upper boundary, as indicated by the arrow. Fig.~\ref{fig:dirfss}a represents the response to a 7-cycle burst with carrier frequency $f=6.7\,\mathrm{Hz}$---belonging to the $5^{\mathrm{th}}$ mode. The wavefield depicts the total mechanical energy landscape in the lattice at a certain instant of propagation. We can see that the wave is mainly propagating along a direction which coincides with OD, while it is attenuated along the directions corresponding to the partial bandgap (left portion of the domain), in complete agreement with the iso-frequency contour of Fig.~\ref{fig:diruc}b. Fig.~\ref{fig:dirfss}b represents the response of the same structure when the carrier frequency of the burst is $f=8.5\,\mathrm{Hz}$---which belongs to the $7^{\mathrm{th}}$ mode. This wavefield displays an opposite pattern: the energy associated with the wave is now mainly propagating in the left portion of the domain, while the right portion remains de-energized. Again, this result is consistent with the unit cell analysis prediction of Fig.~\ref{fig:diruc}c.

To summarize the results shown in this section, we can state that the selective curling of some cantilevers causes a profound modification of the wave anisotropy patterns. In particular, the availability of cantilevers resonating at different frequencies along different directions causes the appearance of partial bandgaps, which lead to spatially-selective and beamed wave patterns. It is also interesting to point out that the same lattice presents opposite wave beaming characteristics at different frequencies. Due to the reversible nature of the curling, we can switch between different directivity patterns by simply curling other sets of cantilevers, ultimately enabling reconfigurable wave beaming.

\begin{figure*} [!htb]
\centering
\includegraphics[scale=1.38]{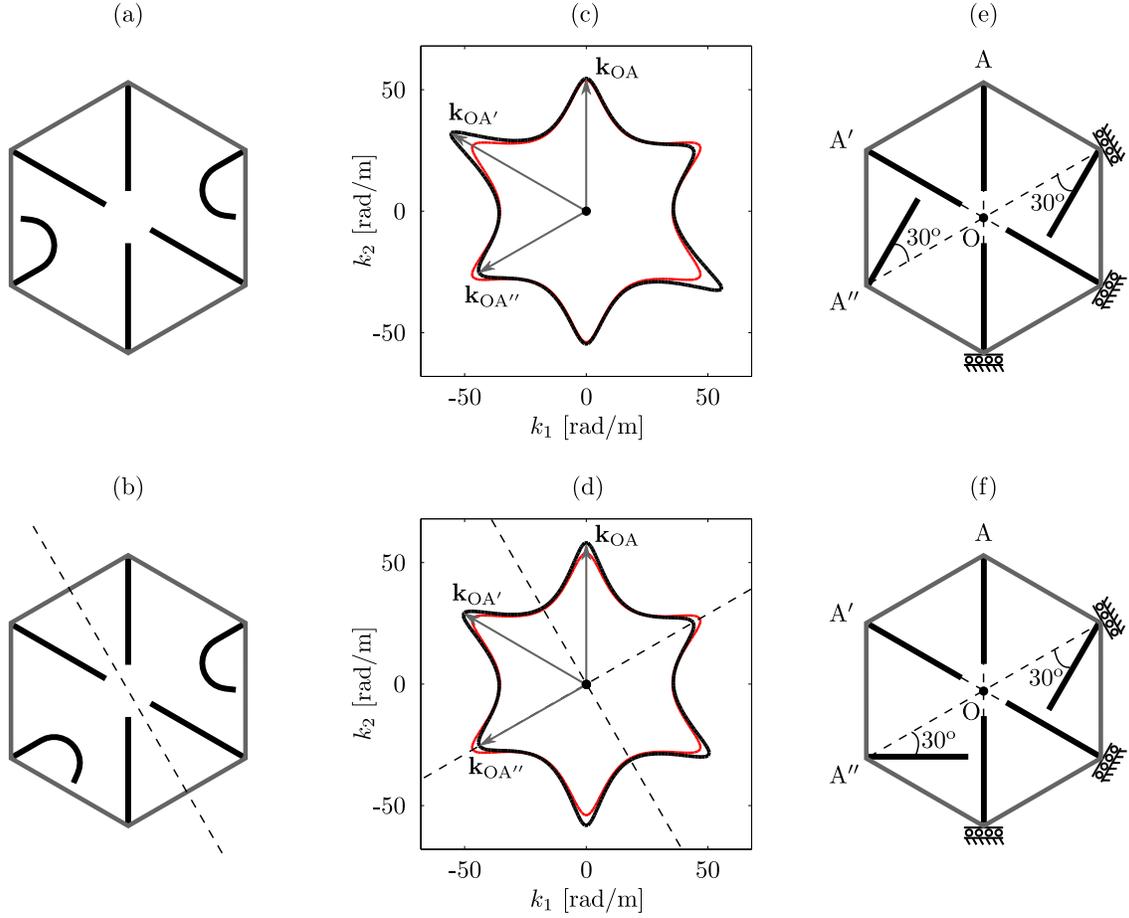}
\caption{Effects of microstructural curling on the symmetry of the wave response: \emph{geometric chirality}. Comparison between two architectures characterized by two curled cantilevers and differing in terms of curling orientation. (a) Geometrically-chiral unit cell (both cantilevers are curled counterclockwise). (b) Non-chiral unit cell (one cantilever is curled clockwise and the other counterclockwise; the dashed line represents an axis of mirror symmetry). (c, d) Responses for the architectures in (a) and (b), respectively (thick black contours; the thin red contours represent the response of an architecture featuring all straight cantilevers). The dashed lines represent mirror axes of symmetry of the response and the arrows indicate wave vectors corresponding to selected directions of propagation. (e, f) Analogous models for the cells in (a) and (b), respectively.}
\label{fig:switch1}
\end{figure*}
\begin{figure*} [!htb]
\centering
\includegraphics[scale=1.38]{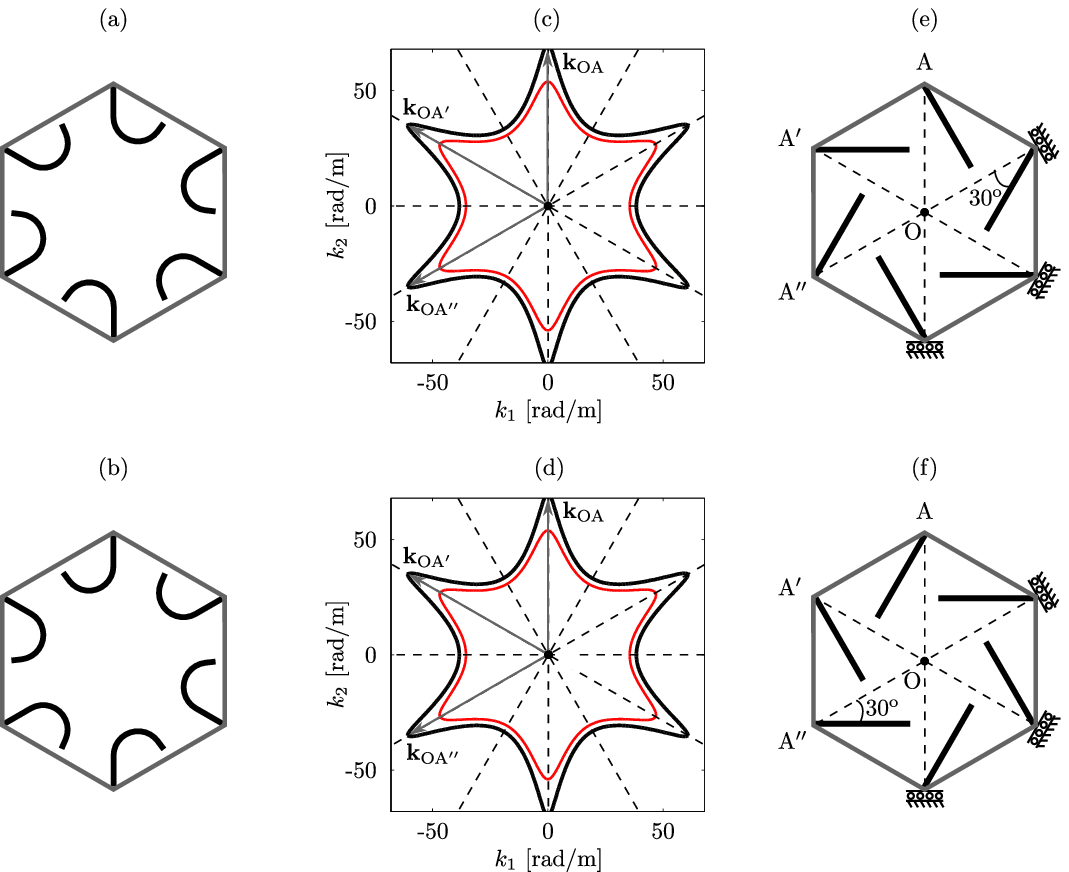}
\caption{Effects of microstructural curling on the symmetry of the wave response: \emph{geometric chirality} versus \emph{functional chirality}. Comparison between two architectures differing in terms of curling orientation of the entire cantilever microstructure. (a, b) Two geometrically-chiral unit cells in which all cantilevers are curled counterclockwise and clockwise, respectively. (c, d) Responses for the architectures in (a) and (b), respectively. The dashed lines represent mirror axes of symmetry of the response and the arrows indicate wave vectors corresponding to selected directions of propagation. (e, f) Analogous models for the cells in (a) and (b), respectively.}
\label{fig:switch2}
\end{figure*}

\subsection{Switchable chirality}
In the previous section, we showed how relaxing the unit cell symmetry has drastic repercussions on the morphology of the wave patterns. To elucidate the rich opportunities for symmetry relaxation attainable with actively-curlable microstructures, we now proceed to provide a mechanistic rationale to link symmetry in the structure to symmetry in the response. For brevity, we restrict our analysis to the S-mode. This choice is motivated by two considerations. First, the deformation patterns associated with the S-mode are particularly easy to interpret, due to the fact that the mode mostly involves flexural deformation of the lattice links~\cite{Celli_JAP_2014}. Secondly, since an S-like mode is observed at low frequencies for virtually every cell configuration, it offers a fair metric of comparison between different architectures.

Let us consider two cell configurations in which only two cantilevers (located at opposite nodes of an hexagonal cell) are curled: in the case of Fig.~\ref{fig:switch1}a, both cantilevers are curled counterclockwise, while in Fig.~\ref{fig:switch1}b one is curled clockwise and the other counterclockwise. By inspecting the symmetry of the cell in Fig.~\ref{fig:switch1}a, we note that the cell is characterized by \emph{geometric chirality}---its shape cannot be recovered, after mirroring it about any axis, by resorting to simple translations and rotations (simply put, the architecture does not possess mirror symmetries). In contrast, the architecture in Fig.~\ref{fig:switch1}b is not chiral, since it features a mirror symmetry about the dashed axis. It is important to realize that the chirality is here introduced through shape modifications of the \emph{auxiliary} microstructure, and can be \emph{switched on/off} through the application of an electric stimulus, without modifying the primary lattice network. In this respect, it is qualitatively different from the chirality most commonly observed in lattice materials, which is associated with a special connectivity of the primary lattice \cite{Spadoni_WM_2009, Liebold_AEM_2013, Trainiti_IJSS_2016}. To emphasize its sole dependency on the microstructure, we refer to it as \emph{second-order chirality}. 

We consider the iso-frequency contour of the S-mode evaluated at $0.49\,f_{max}^{\mathrm{S}}$ (where $f_{max}^{\mathrm{S}}$ is the maximum frequency of the S-mode for any specific architecture). In Fig.~\ref{fig:switch1}c, we report the iso-frequency contour for the case in Fig.~\ref{fig:switch1}a (thick black line), compared to that of a reference case with all straight cantilevers (thin red line), which is characterized by the highest achievable degree of symmetry (6-fold rotational symmetry, 3 mirror axes and inversion symmetry). We observe that the geometrically chiral pattern of Fig.~\ref{fig:switch1}a induces chirality in the response---as highlighted by the lack of mirror symmetries in the black iso-frequency contour in Fig.~\ref{fig:switch1}c. On the other hand, the response of the non-chiral geometry in Fig.~\ref{fig:switch1}b, represented by the black contour in Fig.~\ref{fig:switch1}d, is, as expected, non-chiral---as indicated by the existence of two mirror axes. Alone, these observations would lead to the partial (and overly simplistic) conclusion that geometric chirality in the unit cell implies chirality in the wave response. In the following, we will show that the connection between cell geometry and response, in terms of chirality, is significantly more subtle.

To better elucidate the onset of chirality in the response, we analyze how a wave impinging on the hexagonal cell along three characteristic directions interacts with differently-oriented cantilevers. To provide a more intuitive rationale, we replace the original cell with an analogous one in which the curled cantilevers are substituted with slanted (yet straight) ones, whose centers of mass, just like in the curled case, are off-centered with respect to lines connecting the hexagon's vertices (Fig.~\ref{fig:switch1}e). Note that this analogous model, albeit structurally different, is, for all intents and purposes, identical to the original one in terms of symmetry and geometric chirality, thus providing some useful qualitative information on the wave/cantilever interaction. The three directions of wave propagation we consider are marked as OA, OA$'$ and OA$''$ in Fig.~\ref{fig:switch1}e. These directions correspond to wave vectors ${\mathbf k}_{\mathrm{OA}}$, ${\mathbf k}_{\mathrm{OA}'}$ and ${\mathbf k}_{\mathrm{OA}''}$ in Fig.~\ref{fig:switch1}c and to the inflection points of the iso-frequency contours (where the response chirality manifests the most). When a shear wave impinges on the cell along a direction identified by one of those wave vectors, the unit cell will locally vibrate along a direction perpendicular to the wave vector. In Fig.~\ref{fig:switch1}e, this is schematically denoted by sliding clamp constraints which, individually, only allow translation perpendicularly to the direction of the incoming wave. While the unit cell features a fixed set of internal beam-like resonators, they naturally display different vibrational characteristics according to the way in which they are excited. Our objective is to determine the landscape of \emph{effective} internal resonating mechanisms that is available for waves traveling along different directions. For example, a shear wave impinging along OA (and shaking the cell along the direction perpendicular to OA), engages a cell characterized by two cantilevers parallel to OA (which are activated flexurally), two inclined by 30$^{\mathrm{o}}$ and two inclined by 60$^{\mathrm{o}}$ with respect to OA (these last four will undergo a blend of flexural and axial deformation). A wave along OA$'$, on the other hand, effectively sees a cell in which two cantilevers are inclined by 90$^{\mathrm{o}}$ with respect to OA$'$ (thus activated axially), two are parallel to OA$'$ (thus activated flexurally) and two are inclined by 60$^{\mathrm{o}}$ with respect to OA$'$ (mixed mode). Finally, with respect to a wave along OA$''$, two cantilevers are inclined by 30$^{\mathrm{o}}$ and four are inclined by 60$^{\mathrm{o}}$. In light of these considerations, we can conclude that the establishment of chirality in the response is linked to the availability of three \emph{distinct} sets of resonating mechanisms along the three considered directions. In Fig.~\ref{fig:switch1}f, we repeat the exercise for the architecture in Fig.~\ref{fig:switch1}b. In this case, shear waves along OA and OA$'$ see a cell characterized by an \emph{identical} set of effective resonators, consisting of two cantilevers parallel and one perpendicular to OA (or OA$'$, respectively), two inclined by 60$^{\mathrm{o}}$ and one by 30$^{\mathrm{o}}$ with respect to OA (or OA$'$, respectively). On the contrary, a wave along OA$''$ engages a cell with two cantilevers inclined by 30$^{\mathrm{o}}$ and four inclined by 60$^{\mathrm{o}}$ with respect to OA$''$. Consistently with this additional symmetry in the resonating mechanisms, the iso-frequency contour in Fig.~\ref{fig:switch1}d is identical along OA and OA$'$ and does not display chirality.

Let us now dig deeper into the role of the microstructural elements, to illustrate further implications of the second order chirality. First, we consider the unit cell configuration in Fig.~\ref{fig:switch2}a, characterized by six counterclockwise-curled cantilevers and displaying geometric chirality, as highlighted by the absence of mirror symmetries. Interestingly, and counter-intuitively, its S-mode response, shown in Fig.~\ref{fig:switch2}c (thick black contour), is not chiral (all the dashed lines are axes of mirror symmetry). Identical considerations can be made for the configuration in Fig.~\ref{fig:switch2}b, characterized by six clockwise-curled cantilevers (whose response is shown in Figs.~\ref{fig:switch2}d). These examples suggest that geometric chirality alone does not necessarily imply chirality of the response. To lift this apparent contradiction, we repeat the directional vibration analysis introduced above, here based on the analogous models of Figs.~\ref{fig:switch2}e-f. It is easy to recognize that, in both cases, we have the same identical availability of resonating mechanisms along all directions. This explains why the responses of Figs.~\ref{fig:switch2}c-d are identical. We can argue that these architectures, despite being geometrically chiral, are \emph{functionally non-chiral}---meaning that mirroring the cell about any axis fully preserves the effective \emph{functionality} of the microstructure with respect to the S-mode.

Summarizing our findings, we can conclude that, for lattices with auxiliary microstructures that feature second order geometric chirality, \emph{functional} chirality implies chirality in the response.

\section{Conclusions}
\label{sec:con}
In this work, we have shown that we can resort to the localized shape modification of a population of soft auxiliary microstructural elements to attain a dramatic reconfiguration of the wave characteristics of soft cellular structures. In our structures, the microstructural elements are composite cantilever beams with soft active material inserts, that can curl upon the application of an electric field. This strategy allows for tunable wave control, since the localized curling deformations can be reversed by removing the electric fields. Another remarkable aspect of this strategy is that the wave control capabilities---enabled at the microstructural level---are completely independent from other functionalities and properties of the lattices (e.g., their load-bearing capability). The independent controllability of each cantilever allows considerable flexibility and allows for both spectral and spatial wave control. By curling all the cantilevers inside every unit cell in the same fashion, we can shift the location of the locally resonant bandgap. By curling selected sets of cantilevers in each cell, on the other hand, we relax the symmetry of the architecture, we introduce partial (directional) bandgaps and achieve pronounced wave beaming. Due to the peculiar symmetry landscapes introduced by microstructural curling, we are also able to observe chirality effects of the ``second-order''---i.e., independent from the lattice connectivity and only associated with the mechanical functionality of the microstructures.

\section*{Acknowledgements}
S.G. and P.C. acknowledge the support of the National Science Foundation (grant CMMI-1266089). P.C. also acknowledges the support of the University of Minnesota through the Doctoral Dissertation Fellowship. A.M. and V.T. acknowledge the support of the Air Force Office of Scientific Research (grant FA9550-14-1-0234). Z.O. and S.A. gratefully acknowledge the support of the National Science Foundation (EFRI grant 1240459) and the Air Force Office of Scientific Research. 

\section*{References}
\bibliographystyle{iopart-num}
\balance
\bibliography{myrefs.bib}


\clearpage
\nobalance
\section*{\Large Supplementary Data (SD)}
\renewcommand{\thefigure}{S\arabic{figure}}
\renewcommand{\theequation}{S\arabic{equation}}
\renewcommand{\thepage}{S\arabic{page}}
\renewcommand{\thesection}{S\arabic{section}}
\setcounter{figure}{0}
\setcounter{page}{1}
\setcounter{section}{0}
\setcounter{equation}{0}

\section{Analytical model for nonlinear curling of an electro-actuated cantilever beam}
\label{sec:nlb}
In this section, we describe in detail the model adopted to predict the large deformation of a cantilever beam equipped with patches made of soft active material (an electrostrictive terpolymer, in our case). We begin from a general nonlinear beam formulation, we then specialize it to the case of a cantilever beam in pure bending and, finally, we discuss a strategy to model the electro-actuation.

\subsection{Nonlinear deformation of a shear-indeformable beam}
Our starting point is the framework by Irschik and Gerstmayr [H. Irschik and J. Gerstmayr, \emph{Acta Mech.} {\bf 206}, 1--21, 2009], which provides a formulation for shear-indeformable, nonlinear beams inspired by Reissner's beam theory [E. Reissner, \emph{Z. Angew. Math. Phys.} {\bf 23} 795--804, 1972]. The fundamental assumptions of this formulation are the following:
\begin{itemize}
\item the beam is originally straight;
\item all the fundamental assumptions of Euler-Bernoulli beam theory (cross sections remain undistorted, plane and perpendicular to the neutral axis) hold. These assumptions imply shear-indeformability; thus, implicitly, we are restricting ourselves to the case of very thin beams.
\end{itemize}

\begin{figure} [!htb]
\centering
\includegraphics[scale=1.38]{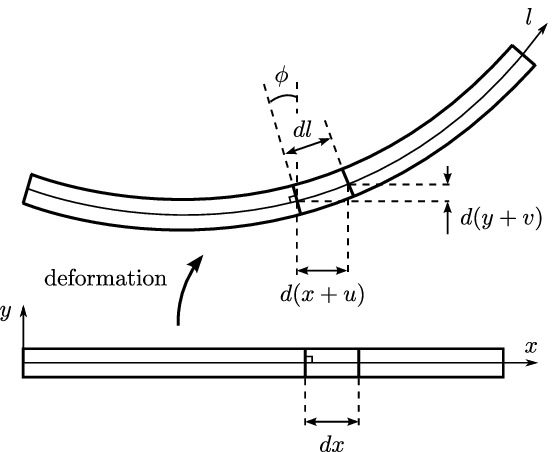}
\caption{Nonlinear deformation of a thin beam.}
\label{fig:f1}
\end{figure}

We consider an initially-straight beam undergoing nonlinear deformation, as sketched in Fig.~\ref{fig:f1}.
In this sketch, $dx$ is the initial length of an infinitesimal beam element, $dl$ is the length of the same infinitesimal element after deformation, $\phi$ is the rotation of the cross section, $u$ is the displacement along the $x$ direction and $v$ is the displacement along the $y$ direction. $dl$ can be rewritten as:
\begin{equation}
dl=\frac{dl}{dx}\,dx=\Lambda_{x0}\,dx=(1+\epsilon_{x0})\,dx\,\,,
\end{equation}
where $\Lambda_{x0}=dl/dx$ is the axial stretch along the centroidal axis and $\epsilon_{x0}=(dl-dx)/dx=\Lambda_{x0}-1$ is the corresponding engineering strain. Note that the only nonzero strain component in this formulation is the axial one. From trigonometry, we can derive the following kinematic relationships:
\begin{equation}
\frac{du}{dx}=(1+\epsilon_{x0})\cos{\phi}-1\,\,,
\label{eq:k1}
\end{equation}
\begin{equation}
\frac{dv}{dx}=(1+\epsilon_{x0})\sin{\phi}\,\,.
\label{eq:k2}
\end{equation}

\begin{figure} [!htb]
\centering
\includegraphics[scale=1.38]{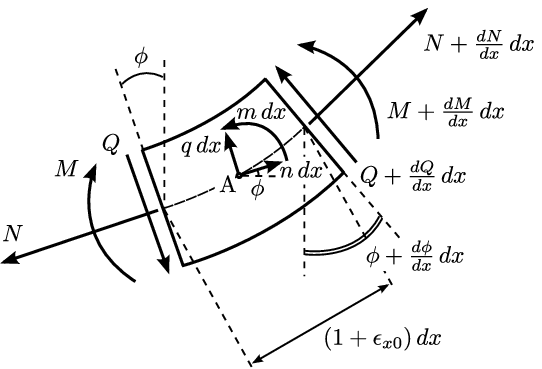}
\caption{Free body diagram of an infinitesimal beam element.}
\label{fig:f2}
\end{figure}

We now shift our attention to the equilibrium equations, which will be initially written for a generic beam and only later specialized to the case of a shear-indeformable beam. We consider a generic (deformed) beam element and we construct the free body diagram shown in Fig.~\ref{fig:f2}.
Here $N$ is the axial force resultant and $n$ is a normal distributed force, $Q$ is the shear force resultant and $q$ is a shear distributed force, $M$ is the moment resultant and $m$ is a distributed moment; the other quantities have been previously introduced. Equilibrium along the horizontal and vertical directions, together with the moment equilibrium about point A, leads to the following equations:
\begin{equation}
N'-Q\,\phi'+n=0\,\,,
\label{eq:eq1}
\end{equation}
\begin{equation}
Q'+N\,\phi'+q=0\,\,,
\label{eq:eq2}
\end{equation}
\begin{equation}
M'+Q\,(1+\epsilon_{x0})+m=0\,\,,
\label{eq:eq3}
\end{equation}
where $(\,\,)'$ stands for $d(\,\,)/dx$. Note that, to obtain these formulae, we assumed $d\phi$ to be small. Next, we eliminate the shear force from the statement of equilibrium. Manipulating Eq.~\ref{eq:eq3}, we obtain the following expression for the shear force:
\begin{equation}
Q=-\frac{M'+m}{1+\epsilon_{x0}}\,\,;
\end{equation}
substituting it into Eq.~\ref{eq:eq1} and Eq.~\ref{eq:eq2}, we obtain the following equations, which enforce the equilibrium of a shear-indeformable beam:
\begin{equation}
N'+\left(\frac{M'+m}{1+\epsilon_{x0}}\right)\,\phi'+n=0\,\,,
\label{eq:eq1f}
\end{equation}
\begin{equation}
N\,\phi'-\left(\frac{M'+m}{1+\epsilon_{x0}}\right)'+q=0\,\,.
\label{eq:eq2f}
\end{equation}

Following [H. Irschik and J. Gerstmayr, \emph{Acta Mech.} {\bf 206}, 1--21, 2009], we can define the constitutive behavior  in terms of nonlinear strain measures and their work conjugates. In particular, we can either introduce a relationship between the Biot stress ($T_{xx}$) and strain ($H_{xx}$), or between the second Piola-Kirchhoff stress ($S_{xx}$) and Green strain ($E_{xx}$), knowing that we can subsequently determine the axial force and moment resultants as:
\begin{equation}
N=\int_{A_0} T_{xx}\, dA=\int_{A_0} \Lambda_x\,S_{xx}\, dA\,\,,
\label{eq:N}
\end{equation}
\begin{equation}
M=-\int_{A_0} T_{xx}\,y\, dA=-\int_{A_0} \Lambda_x\,S_{xx}\,y\,dA\,\,,
\label{eq:M}
\end{equation}
where $A_0$ is the cross-sectional area in the undeformed configuration (which coincides with the deformed area $A$ due to one of the Euler-Bernoulli assumptions), $\Lambda_x=\Lambda_{x0}-y\phi'=1+\epsilon_{x0}-y\phi'$ is the axial stretch ratio at any point of the beam's cross section (a definition which can be derived using continuum mechanics arguments; note that $\phi'=\kappa$ is the generalized curvature of the deformed axis). Note that the expressions for the axial Biot and Green strains in our specific problem are:
\begin{equation}
H_{xx}=\Lambda_x-1=\Lambda_{x0}-y\phi'-1=\epsilon_{x0}-y\phi'\,\,,
\end{equation} 
\begin{equation}
E_{xx}=\frac{1}{2}\left(\Lambda_x^2-1\right)=\frac{1}{2}\left[\left( \Lambda_{x0}-y\phi' \right)^2-1\right]\,\,.
\end{equation} 
It is worth spending few words on the significance of the Biot strain ($H_{xx}$). Recalling from classical small-deformation Euler-Bernoulli theory that the axial engineering strain can be expressed as $\epsilon_x=\epsilon_{x0}-y\kappa$, we notice that, as far as this specific problem is concerned, $H_{xx}\equiv\epsilon_x$. As a consequence, the Biot and Cauchy stresses coincide as well ($T_{xx}\equiv\sigma_x$). The consequences of this fact are significant and will affect the treatment of the electro-actuation: we can base our arguments on engineering strains without falling in the pitfalls of small-deformation analysis.

\subsection{The special case of a cantilever beam undergoing pure bending}
\begin{figure} [!htb]
\centering
\includegraphics[scale=1.38]{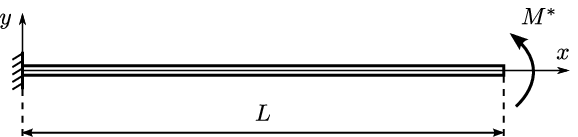}
\caption{Sketch of a cantilever subjected to a tip moment.}
\label{fig:f3}
\end{figure}

The formulation reported in the following is inspired by [A. Muliana, \emph{Int. J. Nonlinear Mech.} {\bf 71}, 152--164, 2015]. In order to model a pure bending scenario, we consider a cantilever beam subjected to a tip moment $M^*$, as sketched in Fig.~\ref{fig:f3}.
Since the beam is undergoing pure bending, the established deformation field features large rotations and small strains. Thus, we can assume a linear elastic constitutive model, i.e., a linear relationship between Biot stress and strain:
\begin{equation}
T_{xx}=E\,H_{xx}=E\,\left( \epsilon_{x0}-y\phi' \right)\,\,.
\label{eq:Txx}
\end{equation}  
Substituting Eq.~\ref{eq:Txx} into Eq.~\ref{eq:N} and Eq.~\ref{eq:M}, we obtain:
\begin{equation}
N=E\,\epsilon_{x0}\int_{A_0} dA- E\,\phi'\int_{A_0} y\,dA\,\,,
\label{eq:Npb}
\end{equation}
\begin{equation}
M=E\,\phi'\int_{A_0} y^2\, dA-E\,\epsilon_{x0}\int_{A_0} y\,dA\,\,.
\label{eq:Mpb}
\end{equation}
If we assume the cross section to be homogeneous, the area is $\int_{A_0} dA=A_0$, the first moment of area is $\int_{A_0} y\,dA=0$ and the second moment of area is $\int_{A_0} y^2\,dA=I$. Thus, we obtain:
\begin{equation}
N=E\,A_0\,\epsilon_{x0}\,\,,
\label{eq:Nh}
\end{equation}
\begin{equation}
M=EI\,\phi'\,\,.
\label{eq:Mh}
\end{equation}
Due to the specific loading and boundary conditions, we have that $N=0$ and $M=M^*$; these equations de facto represent an equilibrium statement and replace Eq.~\ref{eq:eq1f} and Eq.~\ref{eq:eq2f} in this simpler scenario. Substituting $N=0$ and $M=M^*$ into Eq.~\ref{eq:Nh} and Eq.~\ref{eq:Mh}, we obtain:
\begin{equation}
\epsilon_{x0}=0\,\,,
\label{eq:eps}
\end{equation}
\begin{equation}
\phi'=\frac{M^*}{EI}\,\,.
\label{eq:phi}
\end{equation}
As a consequence, we conclude that the neutral axis remains unstretched, while the curvature ($\kappa=\phi'$) is constant throughout the beam's length.

Combining the equilibrium/material equations (Eq.~\ref{eq:eps} and Eq.~\ref{eq:phi}) with the kinematic relationships written in Eq.~\ref{eq:k1} and Eq.~\ref{eq:k2} (which, of course, also hold for the cantilever problem), we can solve for the displacement profile of the cantilever. Integrating Eq.~\ref{eq:phi}, we obtain:
\begin{equation}
\phi=\int_0^x\frac{M^*}{EI}\,dx=\frac{M^*}{EI}\,x\,\,.
\label{eq:phi1}
\end{equation}
Substituting Eq.~\ref{eq:phi1} and Eq.~\ref{eq:eps} into Eq.~\ref{eq:k1} and Eq.~\ref{eq:k2}, and integrating them, we obtain:
\begin{equation}
u(x)=\frac{EI}{M^*}\sin{\left( \frac{M^*}{EI}\,x\right)}-x\,\,,
\end{equation}
\begin{equation}
v(x)=\frac{EI}{M^*}\left[1-\cos{\left( \frac{M^*}{EI}\,x\right)}\right]\,\,.
\end{equation}

\subsection{Electro-actuated smart cantilever}
\begin{figure} [!htb]
\centering
\includegraphics[scale=1.38]{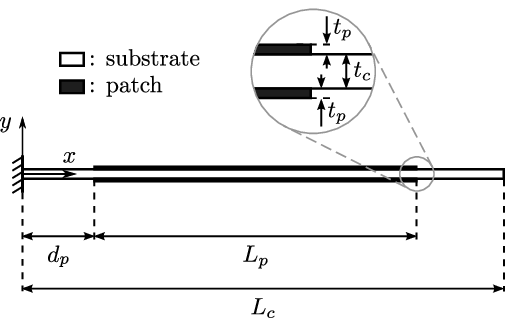}
\caption{Sketch of a cantilever equipped with symmetrically-placed patches (actuator) made of a soft active material. }
\label{fig:f4}
\end{figure}

Our goal is to leverage the shear-indeformable nonlinear beam formulation discussed in the previous sections to predict the curling of a cantilever beam equipped with patches made of a soft active material (in the following, an electro-actuated patch is also referred to as ``actuator''). The sketch of a generic beam configuration is shown in Fig.~\ref{fig:f4}.
The formulation we adopt is largely based on [V. Tajeddini and A. Muliana, \emph{Composite Struct.} {\bf 132}, 1085--1093, 2015], but is here adapted to the particular scenario in which only one patch at a time is activated. In addition to the assumptions introduced in Sec.~\ref{sec:nlb}, we assume the following:
\begin{itemize}
\item the patches are symmetrically-placed with respect to the centroidal axis;
\item the patches are perfectly bonded to the substrate and the thickness of the bonding layer is negligible;
\item the patches are much thinner than the substrate ($t_p\ll t_c$). In light of this, we can consider the axial strain distribution to be constant along the patch's thickness;
\item if the patches were made of a material which was capable of both axial elongation and shrinking, it would have been possible to apply specific electric fields to top and bottom patches in order to achieve a pure bending deformation. However, we are considering patches made of an electrostrictive terpolymer, which can only elongate axially (and not shrink) under the action of an electric field. Thus, we are required to only activate one patch at a time. While this loading configuration does not induce pure bending, we make a quasi-pure bending assumption---as it can be shown that, for the selected parameters, the axial strain is sufficiently small and can be neglected.
\end{itemize}

Due to the pure bending assumption, we consider the action of a patch to be correctly modeled by an equivalent bending moment applied to the beam span where the actuator is located ($d_p<x<d_p+L_p$ in Fig.~\ref{fig:f4}). To evaluate this equivalent moment, we follow an approach developed within ``laminates theory'', in [B-T. Wang and C. Rogers, \emph{J. Intell. Mat. Sys. Struct.} {\bf 2}, 38--58, 1992]. This model, originally developed for small deformations, is here extended to a large-deformation formulation. As a first step, we assume a linear elastic constitutive model for the patch and we express the axial stress in the actuator as:
\begin{equation}
\sigma_{ac}=E_p\,(\epsilon_p-\epsilon_{ac})\,\,,
\label{eq:actstr}
\end{equation}
where $E_p$ is the Young's modulus of the patch, $\epsilon_{ac}$ is the axial strain in the actuator and $\epsilon_p$ is the  actuator free strain, i.e., the strain that an unconstrained actuator would undergo when subjected to an electric field. Note that this relationship written in terms of Cauchy stress ($\sigma$) and engineering strain ($\epsilon$), corresponds to an equivalent one involving Biot stress and strain, since these measures coincide for our specific problem (as discussed in Sec.~\ref{sec:nlb}). Intuitively, due to the constraint placed by bonding, $\epsilon_{ac}<\epsilon_p$. Thus, to have a positive stress when the residual strain $\epsilon_p-\epsilon_{ac}$ is positive, we choose the sign convention as in Eq.~\ref{eq:actstr}.

\begin{figure} [!htb]
\centering
\includegraphics[scale=1.38]{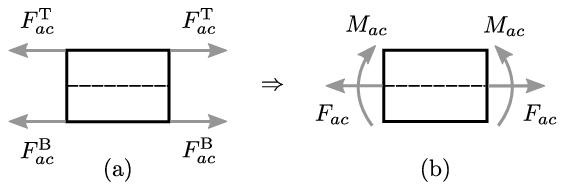}
\caption{(a) Detail of an infinitesimal element of the composite cantilever beam, where the electroactuated patches have been replaced by axial forces applied at the top and bottom surfaces. (b) Same infinitesimal beam element, showing the equivalent axial force and moment that mimick the behavior of the actuators.}
\label{fig:f5}
\end{figure}

We now argue that the thin electroactuated patches can be modeled by axial forces applied at the top and bottom surfaces of the substrate. The forces generated by the two patches are drawn in Fig.~\ref{fig:f5}a.
Depending on their relative magnitude, those forces produce a bending moment, an axial force, or a combination of the two. In order to maximize the bending moment that is achieved, we only activate one of the two patches at a time. This strategy reflects the fact that the terpolymer can only elongate axially. The activation of the bottom (B) patch results in the force $F_{ac}^{\mathrm{B}}=\sigma_{ac}\,b\,t_p$ applied at the bottom surface of the substrate ($y=-t_c/2$), with $F_{ac}^{\mathrm{T}}=0$. This, in turn, produces the following equivalent moment:
\begin{equation}
M_{ac}=F_{ac}^{\mathrm{B}}\,\frac{t_c}{2}=\sigma_{ac}\,b\,\frac{t_p\,t_c}{2}\,\,.
\label{eq:MacB0}
\end{equation}
On the other hand, if the top patch (T) is activated, we have $F_{ac}^{\mathrm{T}}=\sigma_{ac}\,b\,t_p$, $F_{ac}^{\mathrm{B}}=0$ and, consequently, the following equivalent moment:
\begin{equation}
M_{ac}=-F_{ac}^{\mathrm{T}}\,\frac{t_c}{2}=-\sigma_{ac}\,b\,\frac{t_p\,t_c}{2}\,\,,
\label{eq:MacT0}
\end{equation}
where the minus sign is due to the positive moment convention. The equivalent axial force $F_{ac}$, albeit nonzero, is neglected due to the quasi-pure bending assumption.

\begin{figure} [!htb]
\centering
\includegraphics[scale=1.38]{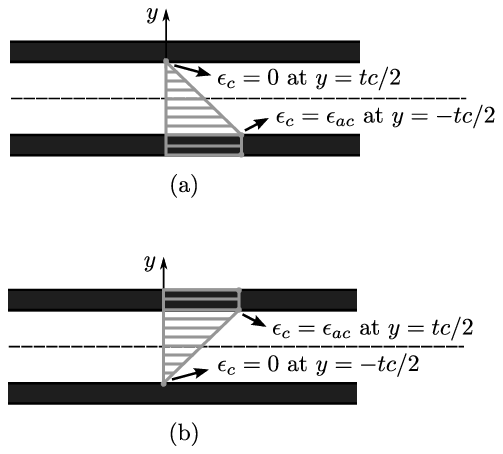}
\caption{Detail of the composite cantilever beam, highlighting the induced strain distribution due to the actuator along the cross section. (a) Case in which the bottom patch (B) is activated. (b) Case in which the top patch (T) is activated.}
\label{fig:f6}
\end{figure}

We now need to determine the moment distribution in the substrate and impose the compatibility of moments at the patch/substrate interface. For the substrate, due to the pure bending assumption, we assume a linear elastic constitutive behavior: $\sigma_c=E\,\epsilon_c$, where $\epsilon_{c}$ is the ``induced strain distribution'' in the substrate due to the actuator. Here $\epsilon_{c}$ is to be interpreted as an induced strain distribution which is compatible with the actuation configuration of the two patches (e.g. if we activate the bottom patch but not the top one, the former will be strained while the latter will not) and that does not represent the actual strain profile in the beam. Following [B-T. Wang and C. Rogers, \emph{J. Intell. Mat. Sys. Struct.} {\bf 2}, 38--58, 1992], we assume this strain to be linearly-distributed, as shown in Fig.~\ref{fig:f6}. Depending on whether we are activating the B or T patch, we face two different scenarios. If we only activate the B patch, we assume the induced strain distribution sketched in Fig.~\ref{fig:f6}a;
this results in $\epsilon_c$ being:
\begin{equation}
\epsilon_c=\epsilon_c^{\mathrm{B}}=\epsilon_{ac}\,\frac{t_c/2-y}{t_c}\,\,.
\label{eq:epscB}
\end{equation}
On the other hand, if we only activate the T patch, the induced strain in the beam is distributed as sketched in Fig.~\ref{fig:f6}b and $\epsilon_c$ can be written as:
\begin{equation}
\epsilon_c=\epsilon_c^{\mathrm{T}}=\epsilon_{ac}\,\frac{t_c/2+y}{t_c}\,\,.
\label{eq:epscT}
\end{equation}
The moment resultant on the substrate due to $\epsilon_c$ (recalling Eq.~\ref{eq:M}) can be computed as:
\begin{equation}
M_c=-\int_{t_c/2}^{tc/2}b\,y\,\sigma_c\,dy=-\int_{t_c/2}^{tc/2}b\,y\,E\,\epsilon_c\,dy\,\,.
\label{eq:M0}
\end{equation}
If we consider the B patch, we substitute the strain distribution in Eq.~\ref{eq:epscB} into Eq.~\ref{eq:M0} and, upon integration, we obtain
\begin{equation}
M_c^{\mathrm{B}}=E\,\frac{b\,t_c^2}{12}\,\epsilon_{ac}\,\,\Rightarrow\,\,\epsilon_{ac}=\frac{12\,M_c^{\mathrm{B}}}{E\,b\,t_c^2}\,\,.
\label{eq:M0B}
\end{equation}
Substituting Eq.~\ref{eq:M0B} into Eq.~\ref{eq:actstr}, plugging the resulting $\sigma_{ac}$ into Eq.~\ref{eq:MacB0} and setting $M_c^{\mathrm{B}}=M_{ac}$ to enforce compatibility, we obtain (after isolating $M_{ac}$ on the left hand side of the equation):
\begin{equation}
M_{ac}=M_{c}^{\mathrm{B}}=\frac{E_p\,E\,b\,t_p\,t_c^2}{2(t_c\,E+6t_p\,E_p)}\,\epsilon_p\,\,.
\label{eq:MacB}
\end{equation}
Similarly, if the T patch is activated, we obtain:
\begin{equation}
M_{ac}=M_{c}^{\mathrm{T}}=-\frac{E_p\,E\,b\,t_p\,t_c^2}{2(t_c\,E+6t_p\,E_p)}\,\epsilon_p\,\,.
\label{eq:MacT}
\end{equation}

At this stage, we still have to model the electromechanical coupling. Recalling that $\epsilon_p$ is the free strain of the patch as a result of the application of an electric field, and invoking a quadratic relation between free strain and electric field (customary for electrostrictive terpolymers), we set:
\begin{equation}
\epsilon_p=\beta\,\left( E^e \right)^2\,\,,
\label{eq:epsp}
\end{equation}
where  $E^e$ is the applied electric field and $\beta$ is the coefficient of electrostriction---which can be experimentally measured. Note that, by substituting Eq.~\ref{eq:epsp} into Eq.~\ref{eq:MacT} and Eq.~\ref{eq:MacB}, we obtain consistent orientations for the beam deformation: if we activate the B patch only, we produce a positive moment, while if we activate the T patch we produce a negative moment.

We now have all the necessary information to derive the displacement profile of a cantilever under the electrostrictive action of one of the two symmetrically-placed actuators. Now recall Eq.~\ref{eq:phi}, rewritten here for this specific problem as:
\begin{equation}
\phi'=\frac{M_{ac}}{EI_c}\,\,,
\label{eq:phin}
\end{equation}
where $M_{ac}$ is the moment modeling the actuator's action and $I_c$ is the second moment of area of the substrate ($I_c=b\,t_c^3/12$). Invoking the kinematic relationships for nonlinear beam theory (Eqs.~\ref{eq:k1}-\ref{eq:k2}) and recalling that $\epsilon_{x0}=0$ for a cantilever in pure bending, we obtain:
\begin{equation}
u'=\cos{\phi}-1\,\,,
\label{eq:k1n}
\end{equation}
\begin{equation}
v'=\sin{\phi}\,\,.
\label{eq:k2n}
\end{equation}
For each beam segment ($0<x<d_p$, $d_p<x<d_p+L_p$, $d_p+L_p<x<L_c$), we need to integrate Eq.~\ref{eq:phin}, substitute it into Eqs.~\ref{eq:k1n}-\ref{eq:k2n}, and integrate again to obtain $u(x)$ and $v(x)$. The final result is the following displacement profile. For $0\leq x\leq d_p$:
\begin{equation}
\left \{\!\!\!\begin{array}{l}
u(x)=0\\
v(x)=0 \end{array}\,\,\,\right.
\label{eq:s1n}
\end{equation}
For $d_p\leq x\leq d_p+L_p$:
\begin{equation}
\left \{\!\!\!\begin{array}{l}
u(x)=\frac{EI_c}{M_{ac}}\sin{\frac{M_{ac}(x-d_p)}{EI_c}} -x+d_p\\
v(x)=\frac{EI_c}{M_{ac}}\left[ 1-\cos{\frac{M_{ac}(x-d_p)}{EI_c}} \right] \end{array}\,\,\,\right.
\label{eq:s2n}
\end{equation}
For $d_p+L_p\leq x\leq L$:
\begin{equation}
\left \{\!\!\!\begin{array}{l}
u(x)\!=\!\cos\!\frac{M_{ac}L_p}{EI_c}(x\!-\!d_p\!-\!L_p)\!-\!x\!+\!d_p\!+\!\frac{EI_c}{M_{ac}}\sin\!{\frac{M_{ac}L_p}{EI_c}}\\
v(x)\!=\!\sin\!\frac{M_{ac}L_p}{EI_c}(x\!-\!d_p\!-\!L_p)\!+\!\frac{EI_c}{M_{ac}}\!\left(1\!-\!\cos\!{\frac{M_{ac}L_p}{EI_c}}\right)\end{array}\right.
\label{eq:s3n}
\end{equation}

\section{Validation of the small-on-large wave model assumption}
In this section, we provide a validation of some of the assumptions invoked to simplify the treatment of wave propagation in soft structures. In particular, our aim is to verify the following:
\begin{enumerate}
\item \label{item1} Can we treat the propagating wave as a small (linear) dynamic perturbation of a structure nonlinearly pre-deformed through the application of a large static load (mimicking the application of an electric field)?
\item \label{item2} Does a fine mesh of straight beam elements accurately describe the behavior of curled beams?
\end{enumerate}
To address these points, we compare some numerical results for a simulation of a curled cantilever beam excited by a time-evolving (burst-like) tip moment. Note that the dimensions of the cantilevers are the same as those reported in Sec.~\ref{sec:curl}, except for the fact that we only consider the PDMS substrate. We approach this simulation in a number of different ways, each invoking an additional layer of simplifying assumptions, to quantify the net effects of these assumptions on the numerical predictions. 

\begin{figure} [!htb]
\centering
\includegraphics[scale=1.38]{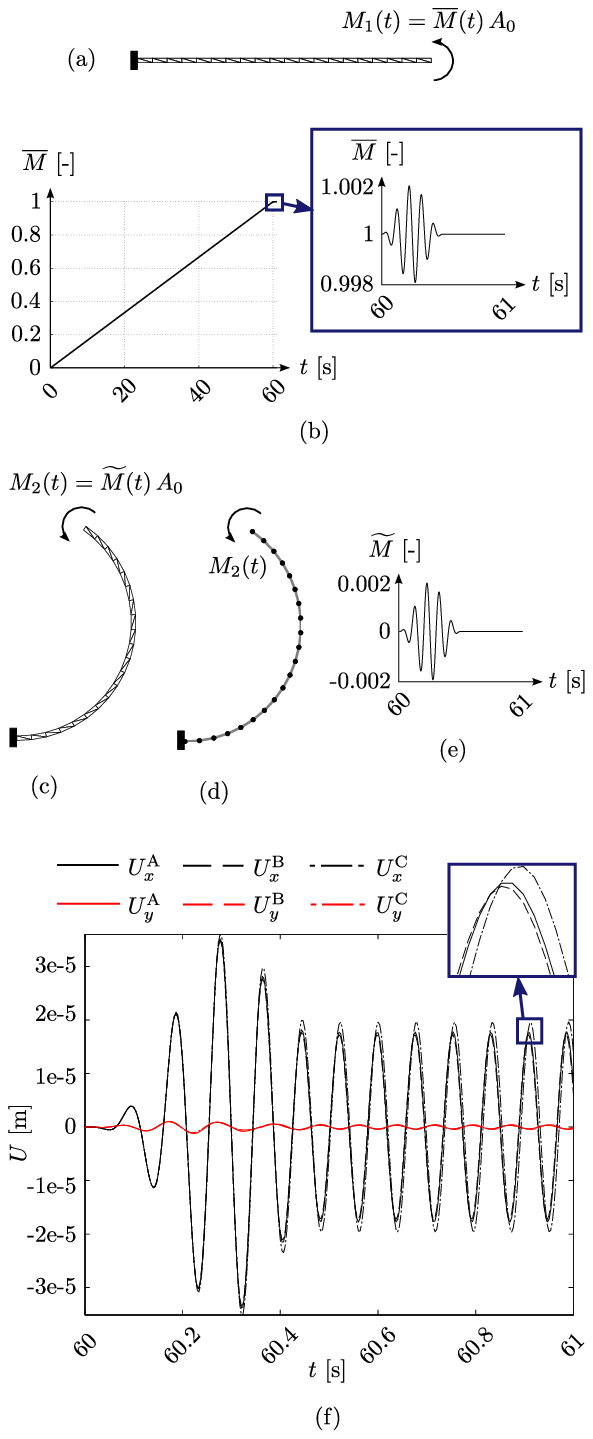}
\caption{Case A: nonlinear transient analysis of a straight beam discretized via isoparametric elements (a), and subjected to a ramp+burst load (b). Case B: nonlinear analysis of a pre-curled beam discretized with isoparametric elements (c). Case C: linear analysis of a pre-curled beam discretized via Timoshenko beam elements (d). (e) Load profile for cases B and C. (f) Results in terms of tip displacements of the beam for cases A, B, C.}
\label{fig:fem}
\end{figure}

First, we resort to a fully nonlinear finite element code to analyze the response of an initially-straight beam, discretized by 6-nodes isoparametric triangular elements (see Fig.~\ref{fig:fem}a), to the excitation profile shown in Fig.~\ref{fig:fem}b. This first scenario is labeled case A.
The load profile comprises an initial slow ramp, during which the beam is brought to its curled state, followed by a burst, which represents a small-displacement perturbation about the curled state. In essence, in this case, all phases of the deformation path are explicitly modeled by the nonlinear code as transient events, although the curling phase, due  to its long time scale, is \emph{de facto} a slow quasi-static event. The amplitude $A_0$ selected for the simulations shown in this section is $A_0=1.54e$-4 [Nm]. Note that, due to the fact that we are dealing with a slender structure subjected to pure moment loading, we expect to induce a small-strain, large-deformation (rotation) field; accordingly, we use a Saint-Venant/Kirchhoff material model---which correctly implements a small-strain linear material law (compatible with the large deformation mechanics framework). Moreover, to guarantee energy and momentum conservation in our model, we use implicit time integration, employing the scheme discussed in the work of Simo and Tarnow [J.C. Simo and N. Tarnow, \emph{Z. Angew. Math. Phys.} {\bf 43}, 757--792, 1992; R. Ganesh, \emph{Mechanisms of wave manipulation in nonlinear periodic structures}, PhD Thesis, University of Minnesota, 2015]. The result of case A, in terms of $x$ and $y$ displacements of the mid-point of the beam's tip, is shown by the solid lines in Fig.~\ref{fig:fem}f.

To address (\ref{item1}), we compare case A to case B---where we considered an already-curled geometry (see Fig.~\ref{fig:fem}c, coinciding with the final state of the ramp portion of the load profile in Fig.~\ref{fig:fem}b) and we apply the load shown in Fig.~\ref{fig:fem}e (which can be obtained from the perturbation part of the load shown in Fig.~\ref{fig:fem}b, by centering it at 0). The result of case B, in terms of $x$ and $y$ displacements of the mid-point of the beam's tip, is shown by the dashed lines in Fig.~\ref{fig:fem}f. We can see that the response is almost identical to that obtained in case A. In light of this, we can conclude that the propagating wave manifests indeed as a perturbation of a pre-deformed state achieved through the application of a large ramp load which is kept constant.

To address (\ref{item2}), we compare the previous results to case C---similar to case B, except for the fact that we consider a linear finite element model comprising Timoshenko beam elements (see Fig.~\ref{fig:fem}d). The result of case C, in terms of $x$ and $y$ displacements of the beam's tip, is shown by the dash-dotted lines in Fig.~\ref{fig:fem}f. We can see that the response of this simplified model agrees qualitatively with cases A and B. The detail of Fig.~\ref{fig:fem}f, however, highlights how the peaks of the response in case C are shifted with respect to A and B. This discrepancy is due to the fact that the dispersion relation is affected by the finite element discretization---since we are using only 20 beam elements per cantilever, we are bound to face frequency-shifts with respect to the true solution. This aspect, which is ininfluential for this proof-of-concept analysis, should be kept in mind when designing experiments to test the strategy presented in this work.

\section{Proper selection of the Brillouin Zone}
\begin{figure} [!htb]
\centering
\includegraphics[scale=1.38]{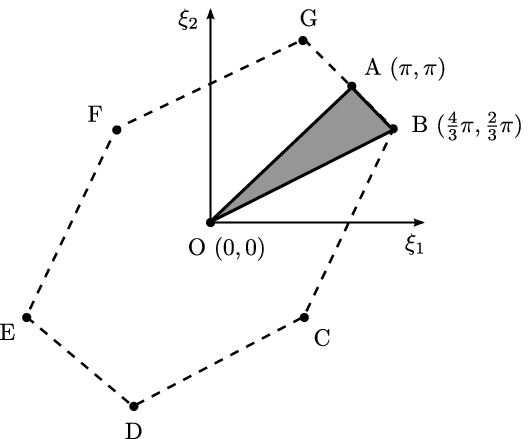}
\caption{First Brillouin zone (BZ, the region bounded by the whole dashed contour) of a regular hexagonal lattice structure and Irreducible Brillouin zone (IBZ, the shaded area) in a cantilever-free case or with cantilevers having all the same properties.}
\label{fig:BZ}
\end{figure}

To access all the information on the wave propagation behavior of a certain periodic structure, it is sufficient to consider the region of the reciprocal wave vector space delimited by the First Brillouin zone (BZ). The BZ for an hexagonal lattice structure is shown in Fig.~\ref{fig:BZ} (the whole region bounded by the dashed contour), where $\xi_1$ and $\xi_2$ are components of the wave vector in the reciprocal lattice coordinate system (see e.g. the work of Gonella and Ruzzene [S. Gonella and M. Ruzzene, J. Sound Vib. 312, 125--139, 2008]).
Note that the BZ only depends on the primary lattice---i.e., it is not affected by the presence of auxiliary microstructural elements, which do not change the lattice connectivity. Due to the symmetries of the unit cell, it is sometimes possible to define an Irreducible Brillouin zone (IBZ): it is then sufficient to investigate wave vectors belonging to this smaller region of the wave vector plane to obtain a full phononic characterization the medium. Since the IBZ strongly depends on the symmetries of the wave response, its shape and size are affected by the presence (and by the curling deformation) of the auxiliary cantilevers: when considering architectures in which the symmetry has been relaxed due to the curling of a subset of cantilevers, the IBZ shown in Fig.~\ref{fig:BZ} (the shaded area) is no longer sufficient. A different IBZ could be identified for a given curling pattern, but it would change as soon as a different internal architecture is established. For these reasons, in order to maintain consistency of the dispersion plots across all cases considered and compared in the manuscript, we decided to consider band diagrams calculated along the contour of the full BZ. Also note that, while in Sec.~\ref{sec:anis} we simply refer to the hexagonal contour as a BZ, this is actually the equivalent representation of the BZ in a Cartesian wave vector space (sometimes known as Wigner-Seitz cell); a Cartesian representation is preferable in studying directivity, as directions in the Cartesian wavevector plane correspond directly to physical directions of wave propagation in the medium.

\begin{figure} [!htb]
\centering
\includegraphics[scale=1.38]{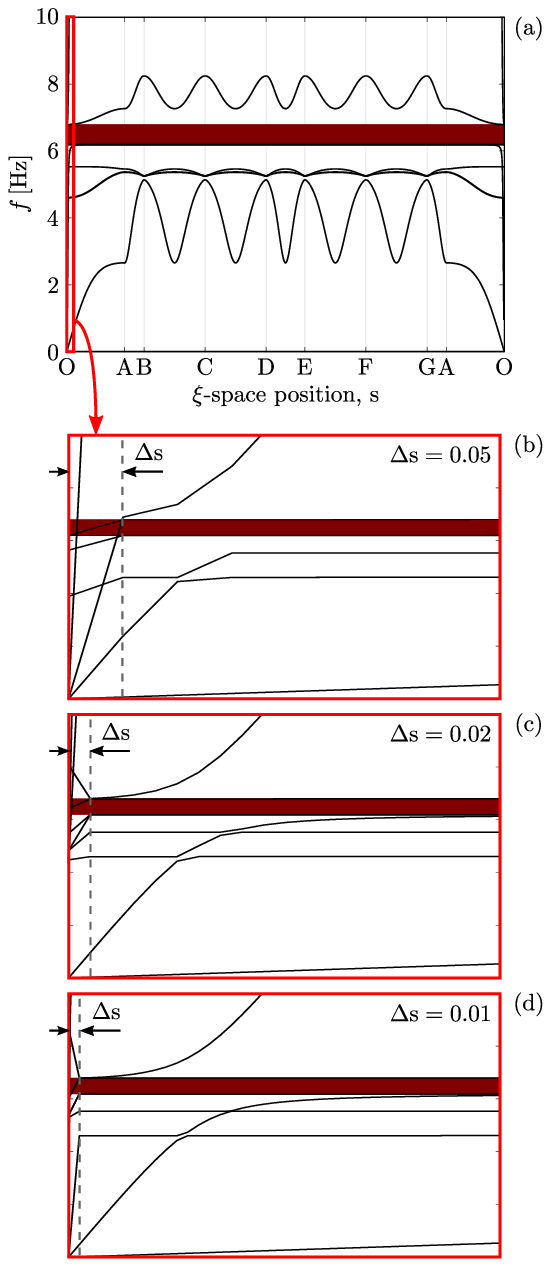}
\caption{Numerical artifacts in the computation of the band diagrams of architectures featuring composite cantilevers. (a) Band diagram for a unit cell featuring all straight cantilevers, where the shaded region indicates the bandgap. The region of interest in this section is highlighted by the red box. (b), (c), (d) Detail of the red box region when $\Delta\mathrm{s}=0.05$, $\Delta\mathrm{s}=0.02$, and $\Delta\mathrm{s}=0.01$, respectively.}
\label{fig:numerr}
\end{figure}
\begin{figure*} [!htb]
\centering
\includegraphics[scale=1.38]{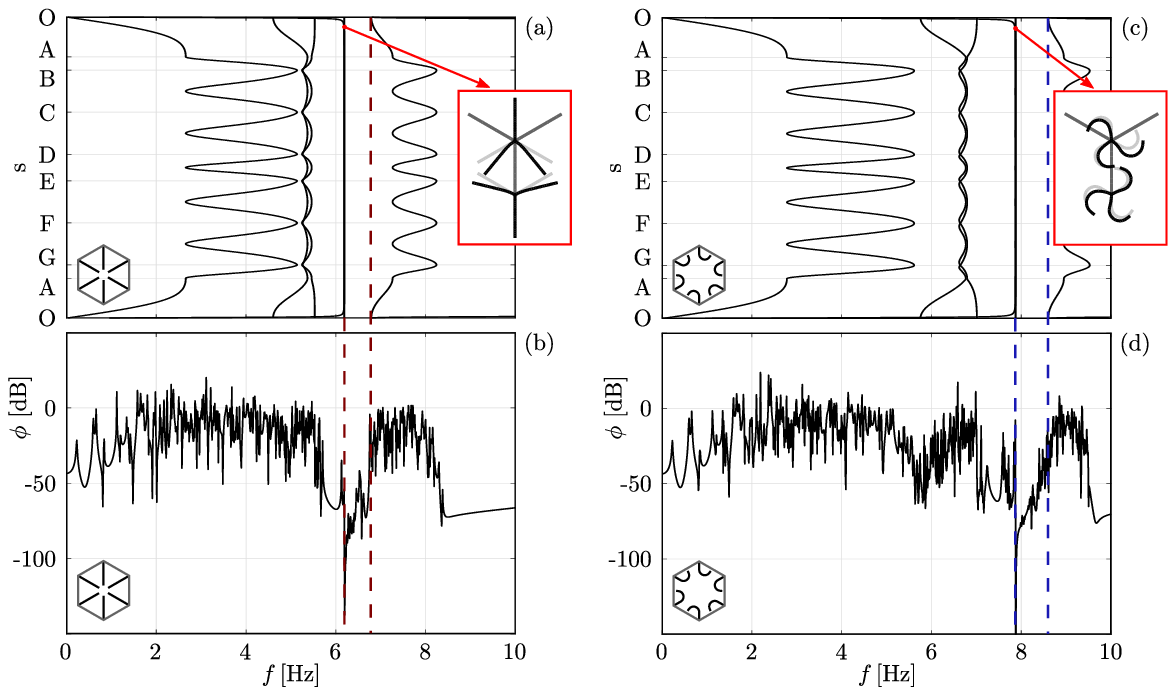}
\caption{(a) Band diagram for a unit cell with straight cantilevers, with detail of the mode shape of the unit cell for a s value along the OA branch and for a mode situated right before the bandgap. In the detail, the light gray, dark gray and black lines are the undeformed cell, the deformed main frame and the deformed cantilevers. (b) Transmissibility of a structure featuring $11\times 10$ cells and all straight cantilevers. (c) Band diagram for a case with curled cantilevers, with detail of a mode shape before the bandgap. (d) Transmissibility of a structure featuring $11\times 10$ cells and all curled cantilevers.}
\label{fig:bandgaps}
\end{figure*}

\section{Numerical error in the band diagram calculation and more bandgap-related information}
In this section, we report on a numerical artifact we encountered when computing the dispersion relation of architectures featuring composite cantilevers. These issues occur only when s (the coordinate spanning the contour of the BZ in the reciprocal wave vector plane---the $\xi$-plane) coincides with the origin of the $\xi$-plane (point O in the BZ), and it manifests through the appearance of non-physical dispersion branches. This issue is illustrated in Fig.~\ref{fig:numerr}. In Fig.~\ref{fig:numerr}a we report the band diagram for a cell configuration in which all cantilevers are straight (non-activated). In order to visualize the numerical artifact, we zoom into the region boxed in red, which extends from $\mathrm{s}=0$ to $\mathrm{s}=0.4$. The portion of the band diagram within this restricted range of s values---for various discretizations of the contour of the BZ (i.e., for different values of $\Delta\mathrm{s}$)---is shown in Figs.~\ref{fig:numerr}b-d. In Fig.~\ref{fig:numerr}b, we can see that the branches behave oddly for s values ranging from 0 to $\Delta\mathrm{s}$, i.e., within the first interval of the discrete s array, in the frequency interval corresponding to the expected bandgap. These strange branches appear mainly around the bandgap region. In order to demonstrate that what we are observing is just a numerical artifact, we look at Figs.~\ref{fig:numerr}c-d and notice that this behavior strongly depends on the discretization of the s vector. Specifically, these branches only exist within the 0 to $\Delta\mathrm{s}$ range for any value of $\Delta\mathrm{s}$ and they seem to change characteristics as we change $\Delta\mathrm{s}$. The numerical nature of these branches is also highlighted by the fact that the error manifests as a high condition number when solving the eigenvalue problem for $\mathrm{s}=0$, while the condition number is reasonably low for all other values of s.

Another piece of evidence that demonstrates the non-physicality of the branches within the first $\Delta\mathrm{s}$ interval is given by the response of finite lattices. In Figs.~\ref{fig:bandgaps}a-b, we report the comparison between the band diagram of a unit cell configuration with straight cantilevers and the transmissibility $\phi$ of a finite lattice comprising $11\times 10$ unit cells and characterized by the same architecture. Note that the transmissibility has been obtained as $\phi=20\,\log (u_{\mathrm{RMS}}^{\mathrm{out}}/u_{\mathrm{RMS}}^{\mathrm{in}})$, where $u_{\mathrm{RMS}}^{\mathrm{out}}$ and $u_{\mathrm{RMS}}^{\mathrm{in}}$ are the root mean squared displacements recorded at a point away from the excitation and near the excitation, respectively (considering the same excitation location and boundary conditions discussed in Sec.~\ref{sec:anis}). The fact that the bandgap is ``through'' and that the branches are a numerical artifact is proven by the appearance of the bandgap in the transmissibility plot. In particular, both onset and end-frequency of the bandgap are in good agreement in both representations of the lattice response. In Fig.~\ref{fig:bandgaps}a, we also report the mode shape for a unit cell at a frequency belonging to a branch occurring immediately before the bandgap. Since the deformed mode shape is superimposed to the undeformed one (light gray line), we can see that the main frame (dark gray line) is undeformed, while the cantilevers are undergoing strong bending-like deformation---an aspect which confirms the locally-resonant nature of these bandgaps [S. Gonella, A.C. To and W.K. Liu, \emph{J. Mech. Phys. Solids} {\bf 57}, 621--633, 2009]. Note that bending mainly occurs near the root of the cantilevers, where the cross section is weaker (featuring PDMS only, without terpolymer layers). The vertical cantilevers are not bent since the point we selected for the computation of the mode shape belongs to the OA segment (this segment, for the P-like waves associated with the considered branch, corresponds to wave vectors associated with wave propagation along the vertical direction, that do not engage the vertical cantilevers). Similar considerations can be made for the architecture featuring curled cantilevers, whose band diagram and transmissibility plots are compared in Figs.~\ref{fig:bandgaps}c-d. Also in this case, the mode shape before the bandgap highlights how the curled cantilevers are bending while the main frame is undeformed. Once again, the bandgap clearly appears in the response of the finite lattice, confirming its presence despite the numerical artifacts in the band diagram.

\end{document}